\RequirePackage{fix-cm}
\documentclass{svjour3}                     
\smartqed  
\usepackage{graphicx}
\journalname{EPJST}
\usepackage{graphicx}%
\usepackage{amsmath,citesupernumber}
\usepackage{amssymb}
\usepackage{dcolumn}
\usepackage{epsfig}
\usepackage{graphics}
\usepackage{bm}

\begin{document}
\newcommand{\al}{\alpha}
\newcommand{\D}{\partial}
\newcommand{\bea}{\begin{eqnarray}}
\newcommand{\eea}{\end{eqnarray}}
\newcommand{\bit}{\begin{itemize}}
\newcommand{\eit}{\end{itemize}}

\title{Generating Finite Dimensional Integrable Nonlinear Dynamical Systems$^*$}

\titlerunning{Generating Finite Dimensional Integrable Nonlinear Dynamical Systems}

\author{M.~Lakshmanan and V.~K.~Chandrasekar}%
\institute{M.~Lakshmanan \at
              Centre for Nonlinear Dynamics, School of Physics,
Bharathidasan University, Tiruchirappalli - 620 024, Tamilnadu, India.
              \email{lakshman.cnld@gmail.com}
\and
 V.~K.~Chandrasekar\at
Centre for Nonlinear Dynamics, School of Physics,
Bharathidasan University, Tiruchirappalli - 620 024, Tamilnadu, India.
\email{chandru25nld@gmail.com}
\and
$^*$Based on the invited lecture delivered by M. Lakshmanan at the NCNSD - 2012}

\maketitle

\date{\today}

\begin{abstract}
In this article, we present a brief overview of some of the recent progress made in identifying and generating finite dimensional integrable nonlinear dynamical systems, exhibiting interesting oscillatory and other solution properties, including quantum aspects. Particularly we concentrate on Lienard type nonlinear oscillators and their generalizations and coupled versions. Specific systems include Mathews-Lakshmanan oscillators, modified Emden equations, isochronous oscillators and generalizations. Nonstandard Lagrangian and Hamiltonian formulations of some of these systems are also briefly touched upon. Nonlocal transformations and linearization aspects are also discussed.


\end{abstract}

\section{Introduction}

Dynamical systems with finite degress of freedom/finite dimensions, whose underlying evolution equations are solvable in terms of analytic functions of elementary and transcendental types, are of extreme importance in physical, enginearing and biological sciences\cite{Guc:1983,Tab:1989,Nayfeh:95,Wig:2003,Lakshmanan:03}. Particularly, systems exhibiting periodic oscillations of different types have considerable applications\cite {Lakshmanan:03,Calogero:08}. Starting from the linear harmonic oscillator equation,
\begin{eqnarray}
\label{cho01}
\ddot{x}+ \omega_0^2x=0 \;\; (^{.}=\frac{d}{dt}),
\end{eqnarray}
with suitable initial conditions, say $x(0)=A$, $\dot{x}(0)=0$, which is solvable in terms of the periodic function
\begin{eqnarray}
\label{cho02}
x(t)=A \cos \omega_0 t,
\end{eqnarray}
one can identify increasing number of autonomous and nonautonomous linear/nonlinear differential equations of different orders, nonlinearity and dimensions which are solvable in terms of suitable functions. Identifying such systems and understanding their solution properties of the underlying dynamical systems and developing new applications in the classical, semiclassical and quantum regimes are of fundamental significance. In this paper, we will be concerned with a class of such nonlinear integrable dynamical systems and their properties.

Consider the cubic undamped, free anharmonic oscillator described by the second order nonlinear ordinary differential equation (ODE),
\begin{eqnarray}
\label{cho03}
\ddot{x}+ \omega_0^2x+\lambda x^3=0 \;\; (^{.}=\frac{d}{dt}),
\end{eqnarray}
with the initial conditions $x(0)=A$, $\dot{x}(0)=0$. The corresponding solution to the initial value problem is
\begin{eqnarray}
\label{cho03a}
x(t)=A cn(\Omega t,k) ,
\end{eqnarray} 
where the frequency of oscillation as a function of the amplitude $A$ is 
\begin{eqnarray}
\label{cho03b}
\Omega=\sqrt{\omega_o^2+\lambda A^2},
\end{eqnarray} 
and the modulus square of the Jacobian elliptic function is
\begin{eqnarray}
\label{cho03c}
k^2=\frac{\lambda A^2}{2(\omega_0^2+\lambda A^2)}.
\end{eqnarray} 
Note that the frequency now depends on the amplitude or initial condition, which is a characteristic feature of typical nonlinear oscillators. In fact this dependence can become highly sensitive when suitable additional external force, nonlinearity, damping, etc. are added, leading to bifurcations and chaos \cite{Nayfeh:95,Lakshmanan:03}. A typical example is the Duffing oscillator\cite{Guc:1983,Tab:1989,Nayfeh:95,Wig:2003,Lakshmanan:03} $\ddot{x}+ \alpha \dot{x}+\omega_0^2x+\lambda x^3=f \cos \omega t$.

Then, the question arises as to whether nonlinear system always admit elliptic or higher functions and whether amplitude dependence of frequency of oscillations is a fundemental property of such oscillators. In this paper we will demonstrate that a large class of interesting nonlinear dynamical systems admitting elementary periodic solutions with or without dependence on initial conditions exist, apart from systems solvable by more complicated functions, and show the possibility of amplitude independent frequency of oscillations (isochronous property\cite {Calogero:08}) and other related properties in a large class of nonlinear systems.

Specifically, we will consider the following class of dynamical systems (the following naming is our personal choice for convenience):\\
(a) Lienard type I:
\begin{eqnarray}
\label{cho04}
\ddot{x}+ f(x)\dot{x}^2+g(x)=0
\end{eqnarray}
(b) Lienard type II:
\begin{eqnarray}
\label{cho05}
\ddot{x}+ h(x)\dot{x}+g(x)=0
\end{eqnarray}
(c) Lienard type III:
\begin{eqnarray}
\label{cho06}
\ddot{x}+ f(x)\dot{x}^2+h(x)\dot{x}+g(x)=0
\end{eqnarray}
(d) Lienard type IV:
\begin{eqnarray}
\label{cho07}
\ddot{x}+ f(x,t)\dot{x}^2+h(x,t)\dot{x}+g(x,t)=0
\end{eqnarray}
(e) Coupled versions of the above Li\'enard type of dynamical systems:
\begin{eqnarray}
\label{cho08}
&&\ddot{x}+ f_1(x,y,t)\dot{x}^2+f_2(x,y,t)\dot{y}^2+f_3(x,y,t)\dot{x}\dot{y}
\nonumber\\&&\qquad\qquad\qquad\qquad+h_1(x,y,t)\dot{x}
+h_2(x,y,t)\dot{y}+g_1(x,y,t)=0\nonumber\\
&&\ddot{y}+ f_4(x,y,t)\dot{x}^2+f_5(x,y,t)\dot{y}^2+f_6(x,y,t)\dot{x}\dot{y}
\nonumber\\&&\qquad\qquad\qquad\qquad+h_3(x,y,t)\dot{x}
+h_4(x,y,t)\dot{y}+g_2(x,y,t)=0.
\end{eqnarray}
Here $f,g,h$, etc. are arbitrary functions in the indicated variables.
Considering Li\'enard type I equation (\ref{cho04}), one can identify an interesting class of nonlinear oscillators with velocity dependent or position dependent mass Hamiltonians of the form
\begin{eqnarray}
\label{cho09}
H=\frac{1}{2}F(x)p^2+V(x),
\end{eqnarray}
having very interesting classical properties and quantum spectrum (see Sec. 2 for more details). 3-dimensional and N-dimensional generalizations corresponding to motion of dynamical systems on 3 and N dimensional spheres can be identified. Underlying Lie point symmetries of (\ref{cho04}) can be classified as linearizable and integrable ones.

On the other hand system (\ref{cho05}), though in general nonintegrable, can admit integrable dynamical systems under certain conditions. In particular, it can admit a class of nonstandard type Hamiltonian systems of the form 
\begin{eqnarray}
\label{cho10}
H=\frac{1}{2}F(p)x^2+U(p),
\end{eqnarray}
and generalizations (see Sec. 3 for further details). Of particular interest here is the modified Emden equation (MEE), which is of {\it PT} symmetric type and can be quantized in momentum space. N-dimensional generalizations of the above type of systems can also be identified. Finally we will consider some of the integrable versions of Eqs. (\ref{cho06})-({\ref{cho08}) also in this paper. 

The plan of the paper is as follows. In Sec. 2, we discuss Li\'enard type nonlinear oscillators (\ref{cho04}) with quadratic velocity terms. In particular, we discuss Mathews-Lakshmanan oscillators and their generalizations and discuss their classical and quantum properties. In Sec. 3, we analyse the Li\'enard type oscillators (\ref{cho05}) with linear velocity terms. Special attention is given to {\it PT} symmetric modified Emden equation (MEE) and generalizations and discuss the associated nonstandard Lagrangian and Hamiltonian formulations. Quantum aspects of the systems are also discussed. Higher dimensional integrable generalizations of MEEs are discussed in Sec. 4, while integrable Li\'enard type systems of the types III-IV are briefly discussed in Sec. 5, including infinite hierarchies. In Sec. 6, we briefly discuss other interesting systems including Painlev\'e and Gambier equations and present a brief outline of further challenges.

\section{Li\'enard Type I systems: Mathews-Lakshmanan oscillators and generalization}
Consider the dynamical equation
\begin{eqnarray}
\label{cho11}
\ddot{x}+ f(x)\dot{x}^2+g(x)=0.
\end{eqnarray}
Multiplying by an integrating factor $\dot{x}e^{2\int f(x)dx}$, one can obtain after one integration\cite{Murphy:69}
\begin{eqnarray}
\label{cho12}
\dot{x}^2e^{2\int f(x)dx}+ 2\int g(x)e^{2\int f(x)dx}=I,
\end{eqnarray}
where $I$ is an integral of motion. In general this leads to velocity dependent\cite{Mathews:74} or position dependent mass \cite{bend:98} Hamiltonian systems, depending on the forms of the functions $f(x)$ and $g(x)$:
\begin{eqnarray}
\label{cho13}
H(x)=F(x)\frac{p^2}{2}+V(x),
\end{eqnarray}
where $p(x)$ is the canonically conjugate momentum, while $F(x)$ and $V(x)$ are functions related to $f(x)$ and $g(x)$.

\subsection{Mathews-Lakshmanan (ML) oscillators}
One of the most interesting examples is the case
\begin{eqnarray}
\label{cho14}
f(x)=\frac{\lambda x}{(1-\lambda x^2)},\;g(x)=\frac{\omega_0^2x}{(1-\lambda x^2)},
\end{eqnarray}
where $\lambda$ and $\omega_0$ are constant parameters. Note that in the above, when $\lambda>0$, the range of displacement is restricted as $-1/\sqrt{\lambda}<x<1/\sqrt{\lambda}$, while for $\lambda<0$, there is no such restriction. The resulting system (\ref{cho11}) is the Mathews-Lakshmanan oscillator\cite{Mathews:74} equation,
\begin{eqnarray}
\label{cho15}
\ddot{x}+ \frac{\lambda x}{(1-\lambda x^2)}\dot{x}^2+\frac{\omega_0^2x}{(1-\lambda x^2)}=0,
\end{eqnarray}
leading to the nonpolynomial Hamiltonian/Lagrangian
\begin{eqnarray}
\label{cho16}
H=\frac{1}{2}\bigg[p^2(1-\lambda x^2)+\frac{\omega_0^2x^2}{(1-\lambda x^2)}\bigg],\;\;\;\;\;\;L=\frac{1}{2}\bigg[\frac{\dot{x}^2-\omega_0^2x^2}{(1-\lambda x^2)}\bigg],
\end{eqnarray}
where
\begin{eqnarray}
\label{cho17}
p=\frac{\dot{x}}{(1-\lambda x^2)}.
\end{eqnarray}
The ML oscillator equation (\ref{cho15})  may be considered as the zero-dimensional version of a scalar nonpolynomial field equation\cite{Delbourgo} or as a velocity dependent potential oscillator.  It can also be considered as an oscillator with a position dependent effective mass\cite{Koc}.
The nonpolynomial oscillator system (\ref{cho15}) exihibits simple harmonic periodic solutions but with amplitude dependent frequency,
\begin{eqnarray}
\label{cho18}
x(t)=A\cos(\Omega t+\delta), \qquad\Omega=\frac{\omega_0}{\sqrt{1-\lambda A^2}}.
\end{eqnarray}
The phase plane structure for the cases $\lambda>0$ and $\lambda<0$ are shown in Fig. 1.

It is interesting to note that the system (\ref{cho15}) is linearizable under the nonlocal transformation\cite{Gladwin Pradeep:09a}
\begin{eqnarray}
\label{cho19}
U=\frac{x}{\sqrt{1-\lambda x^2}},\;\; d\tau=\frac{1}{(1-\lambda x^2)}dt,
\end{eqnarray}
so that
\begin{eqnarray}
\label{cho20}
\frac{d^2U}{d\tau^2}+\omega_0^2U=0.
\end{eqnarray}
Solving (\ref{cho20}) and using (\ref{cho19}), one can indeed recover back the solution (\ref{cho18}).

The quantum version of (\ref{cho16}) is also exactly solvable\cite{Mathews:75}. Symmetrizing the classical Hamiltonian in the quantum case as 
\begin{eqnarray}
\label{cho21}
H_q=\frac{1}{2}\bigg[\frac{1}{2}\hat{p}^2(1-\lambda \hat{x}^2)+\frac{1}{2}(1-\lambda \hat{x}^2)\hat{p}^2+\frac{\omega_0^2\hat{x}^2}{(1-\lambda \hat{x}^2)}\bigg],
\end{eqnarray}
(where hat stands for linear differential operators) the time-independent Schr\"{o}dinger equation $H_q\psi=E\psi$ can be written as the linear eigen value problem
\begin{eqnarray}
\label{cho22}
(1-\lambda x^2)\frac{d^2\psi}{dx^2}-2\lambda x \frac{d\psi}{dx}+\bigg[\frac{2E}{\hbar^2}+
\frac{\omega_0^2}{\lambda \hbar^2}-\lambda-\frac{\omega_0^2}{\lambda \hbar^2}(\frac{1}{(1-\lambda x^2)}\bigg]\psi=0.
\end{eqnarray}

On solving (\ref{cho22}), for appropriate boundary conditions, the energy spectrum and eigenfunctions turn out to be the following:
(i) $\lambda>0$:
\begin{eqnarray}
\label{cho23}
E_n=(n+\frac{1}{2})\hbar \omega_0+\frac{1}{2}\lambda \hbar^2 (n^2+n+1),\;\;n=0,1,2,\ldots
\end{eqnarray}
and 
\begin{equation}
\psi_n = \left\{\begin{array}{cc} N_n P_{n+\mu}^{-\mu}\bigg(\lambda^{1/2}x\bigg),
              \quad |x|<\lambda^{-1/2}, nonumber \\
                                  \hspace{-4cm} \qquad\qquad 0, \quad |x|>\lambda^{-1/2},\;\mu=\frac{\omega_0}{\lambda \hbar}.
                                   \end{array}\right.,
\label{cho24}
\end{equation}
(ii) $\lambda<0$:\\
(a) Bound States:
\begin{eqnarray}
\label{cho25}
E_n=(n+\frac{1}{2})\hbar \omega_0-\frac{1}{2}|\lambda| \hbar^2 (n^2+n+1),\;\;n=0,1,2,\ldots, N.
\end{eqnarray}
and 
\begin{equation}
\psi_n = \left\{\begin{array}{cc} N_n Q_{\mu-(n+1)}^{\mu}\bigg(|\lambda|^{1/2}x\bigg),
              \quad |x|<|\lambda|^{1/2}, \nonumber\\
                                  \hspace{-4cm} \qquad\quad 0, \quad |x|>|\lambda|^{1/2}.
                                   \end{array}\right.,
\label{cho24a}
\end{equation}
(b) Scattering states:
\begin{eqnarray}
\label{cho27}
\psi(x)=NP_{-\frac{1}{2}+ip}^{-\mu}\bigg(\lambda^{1/2}x\bigg)\Theta\bigg(|\lambda|^{1/2}x-1\bigg).
\end{eqnarray}
In the above $P_{\mu}^{\gamma}(x)$and $Q_{\mu}^{\gamma}(x)$ are associated Legendre functions\cite{Mathews:75} and $\Theta$ is the Heaviside step function.
\subsection{Three-dimensional and N-dimensional generalizations:}
A three dimensional generalization of the ML oscillator (\ref{cho15}) corresponds to the Hamiltonian\cite{Lakshmanan:75}
\begin{eqnarray}
\label{cho28}
H=\frac{1}{2}\bigg[\vec{p}^2-\lambda(\vec{p}.\vec{q})^2+\frac{\omega_0^2\vec{q}^2}{(1-\lambda\vec{q}^2)}\bigg],\;\;\vec{q}=(q_1,q_2,q_3).
\end{eqnarray}
It is the zero-dimensional isoscalar version of the SU(2)$\otimes$SU(2) chiral model in the Gasiorowicz-Geffen coordinates\cite{Lakshmanan:75}.  The associated canonical equations of motion can be rewritten in the form
\begin{eqnarray}
\label{cho29}
\ddot{q}_i+\left[\frac{\lambda(\vec{q}.\ddot{\vec{q}})}{(1-\lambda\vec{q}^2)}+\frac{\lambda\dot{\vec{q}}^2}{(1-\lambda\vec{q}^2)}+\frac{\lambda^2(\vec{q}.\dot{\vec{q}})^2}{(1-\lambda\vec{q}^2)^2}+\frac{\omega_0^2}{(1-\lambda\vec{q})^2}\right]q_i=0.
\end{eqnarray}
In spherical polar coordinates
\begin{eqnarray}
\label{cho30}
\vec{q}=(q\sin\theta\cos\phi,q\sin\theta\sin\phi,q\cos\theta),
\end{eqnarray}
Eq. (\ref{cho29}) can be separated out as 
\begin{eqnarray}
\label{cho31}
&&\dot{\phi}=\frac{C_1}{q^2\sin^2\theta},\;\;\quad q^4\dot{\theta}^2=C_2^2-\frac{C_1^2}{\sin^2\theta},\nonumber\\
&&\ddot{q}+\frac{\lambda q \dot{q}^2}{1-\lambda q^2}+\frac{\omega_0^2 q}{1-\lambda q^2}=\frac{C_2^2(1-\lambda q^2)}{q^3},
\end{eqnarray}
where $C_1$ and $C_2$ are constants.
On integration one can write down the solution as
\begin{eqnarray}
\label{cho32}
 q(t)=A\left[1-\beta\sin^2\Omega t\right]^{\frac{1}{2}},\quad\Omega^2=\frac{\omega_0^2}{(1-\lambda A^2)}+\frac{\lambda C_2^2}{A^2}
\end{eqnarray}
where $\beta=1-\frac{1}{\lambda A^2}\bigg(1-\lambda A^2-\frac{\omega_0^2-\lambda^2 C_2}{\Omega^2}\bigg)$ and $A$ is the  integration constant.

In fact the above system can be interpreted\cite{higgs} as an isotropic oscillator moving on a 3-sphere, $q_1^2+q_2^2+q_3^2=\sqrt{\lambda}$.

The associated quantum Hamiltonian can be symmetrized in the form
\begin{eqnarray}
\label{cho33}
H_q=\frac{1}{2}\vec{\hat{p}}^2-\lambda(\vec{\hat{p}}.\vec{\hat{q}})(\vec{\hat{q}}.\vec{\hat{p}})+\frac{\omega_0^2\vec{\hat{q}}^2}{1-\lambda \vec{\hat{q}}^2}.
\end{eqnarray}
From the Lie symmetries associated with system (\ref{cho29}), one can identify the following symmetry generators and Lie algebra for the quantum system (\ref{cho33}):
\begin{eqnarray}
\label{cho34}
&&J_i=\epsilon_{ijk}(q_jp_k-p_jq_k),\;\;\;i,j,k=1,2,3,\nonumber\\
&&F_i=\frac{1}{2}\left[(1-\lambda\vec{q}^2)^{\frac{1}{2}}p_i+p_i(1-\lambda \vec{q}^2)^{\frac{1}{2}}\right],
\end{eqnarray}
so  that a nonlinear algebra can be realized.
\begin{eqnarray}
\label{cho35}
&&[J_i,J_j]=i\epsilon_{ijk}J_k,\,\,\,\,\,[F_i,F_j]=i\lambda\epsilon_{ijk}J_k,\,\,\,
[J_i,F_j]=\frac{i}{\lambda}\epsilon_{ijk}F_k,
\nonumber\\
&&[F_i,q_j]=-\frac{1}{\lambda}\sqrt{1-\lambda q^2}\delta_{ij}.
\end{eqnarray}
Depending on whether $\lambda>0$ or $\lambda<0$, one has an SO(4) or SO(3,1) algebra, respectively. The problem can be generalized to N-dimensionas and the associated nonlinear algebra can be written down as was done by Higgs\cite{higgs} and Leemon\cite{Leemon:79}. Such algebras in fact are the precursors to the study of quantum groups\cite{13}.

\begin{figure*}
\includegraphics[width=12.0cm,height=6.90cm]{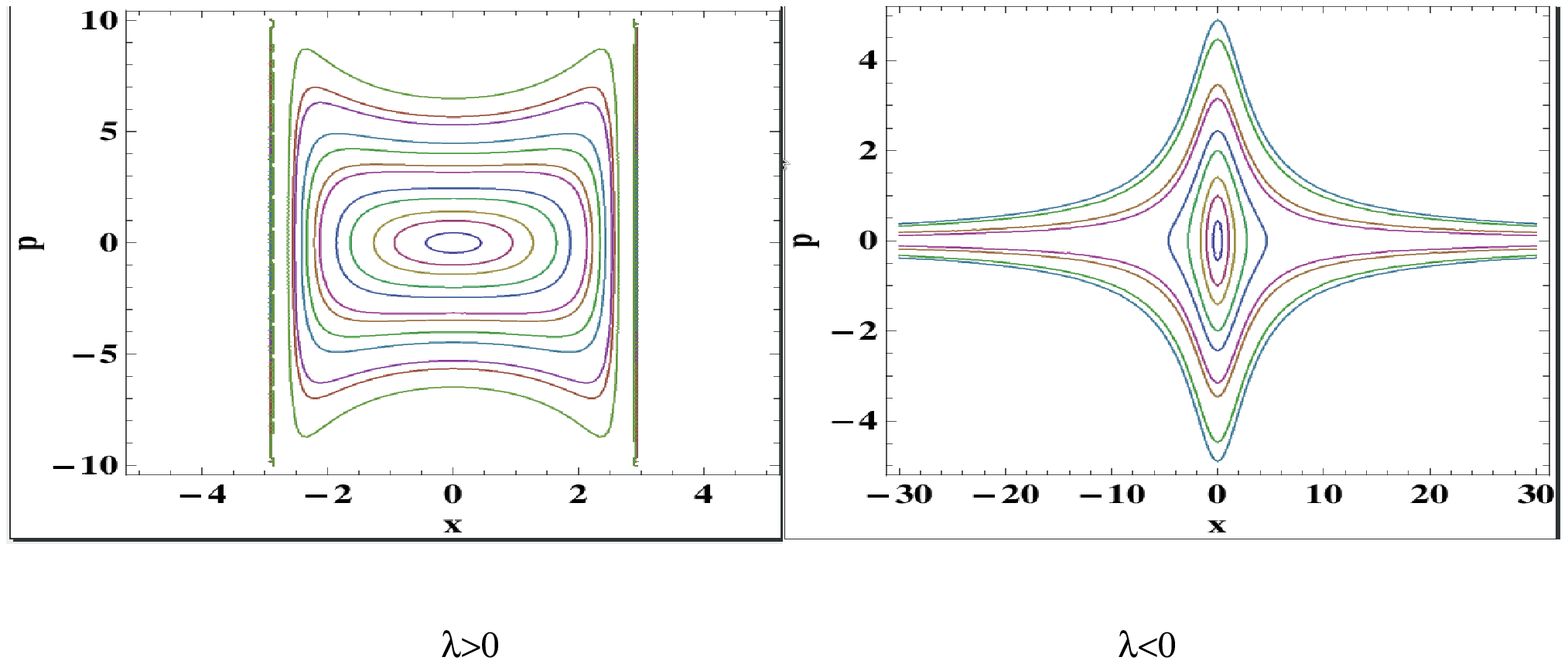}
\caption{The phase plane structure of Mathews-Lakshmanan oscillator (\ref{cho15}) (left) $\lambda>0$ and (right) $\lambda<0$.}
\label{fig1}
\end{figure*}

The three dimensional quandum problem itself can be solved completely\cite{Lakshmanan:75} namely $H_q\psi_{nlm}=E_n\psi_{nlm}$, in spherical polar coordinates. The eigenvalues and eigenfunctions turn out to be
\begin{eqnarray}
\label{cho36}
&&E_n=(n+\frac{3}{2}\hbar\omega_0)+\frac{1}{2}\lambda\hbar^2(n^2+3n),(n=2n_i+l),\;i=1,2, \ldots\nonumber\\
&&\psi_n=A(1-\lambda q^2)^{\frac{1}{2}}q^{\mu-1}F(\frac{1}{2}\mu+\frac{1}{2}\nu+\frac{1}{2}p,-n;\mu+\frac{1}{2},\lambda q^2)\nonumber\\
&&\hspace{2cm}\times Y_{lm}(\theta,\phi),\quad 0\le q<\lambda^{-\frac{1}{2}}
\end{eqnarray}
where $Y_{lm}$ are the spherical harmonics and $F(\alpha,\beta,\gamma;x)$ is the hypergeometric function.
For certain class of N-dimensional version of the above systems corresponding to an isotropic oscillator moving on an N-sphere, one can again solve the quantum machanical eigenvalue problem. For a complete solution, see for example\cite{higgs}. These problems continue to evoke considerable interest in the literature in their classical and quantum versions for their oscillatory properties, energy level spectra, eigenfunctions, construction of creation and annihilation operators, coherent states and generalizations leading to bifurcations and chaos\cite{15,Carinena:04,Carinena:07,Tewari:13,Bhuneshwari:12,Bagchi:13,Cruz:13}.

A particularly interesting completely integrable N-degrees of freedom system introduced by Carinena et al.\cite{Carinena:04} is
\begin{eqnarray}
\label{cho37}
H=\frac{1}{2}\left[\sum p_i^2+\lambda(\sum_{i=1}^{N} q_ip_i)^2\right]+\frac{\alpha^2}{2}\left(\frac{q^2}{1-\lambda q^2}\right),\,\,q^2=\sum_{i=1}^N q_i^2.
\end{eqnarray}
The higher dimensional equation (\ref{cho37}) is superintegrable with $2N-1$ quadratic constants of motion. 

Finally the case 
\begin{eqnarray}
\label{cho37a}
f(x)=\frac{\lambda x}{(1+\lambda x^2)},\;\;\; g(x)=\frac{\omega_0^2}{(1+\lambda x^2)}
\end{eqnarray}
in Eq. (\ref{cho11}) is also of considerable interest which corresponds to the motion of a particle on a rotating parabola and the corresponding damped and forced version exhibits a rich variety of bifurcations and chaos phenomena, see for more details ref.\cite{15,Bagchi:13}.
\subsection{Lie point symmetries and isochronous oscillators}
It is of interest to consider the Lie point symmetries\cite{Tewari:13} associated with Eq. (\ref{cho04}). It can be shown that it admits eight, three, two and one parameter symmetries. In particular, the case corresponding to eight parameter symmetries is linearizable under coordinate transformations, in confirmty with the general theory of Lie on group classification of ODEs. This case corresponds to the choice 
\begin{eqnarray}
\label{cho38}
g(x)=e^{-\int f(x)dx}\left[g_1\int dx e^{\int f(x)dx}+g_2\right],
\end{eqnarray}
where $g_1$ and $g_2$ are arbitrary constants and $f(x)$ is any given function in Eq. (\ref{cho11}). In this case, one can consider a transformation of the form 
\begin{eqnarray}
\label{cho39}
X(x)=\frac{g_1}{\omega_0^2}\left[\int e^{\int f(x)dx}dx+\frac{g_2}{\omega_0^2}\right]
\end{eqnarray}
which linearizes equation (\ref{cho04}) to the linear harmonic oscillator form (\ref{cho01}). Consequently the solution of (\ref{cho04}) will be periodic with period $T=\frac{2\pi}{\omega_0}$, which is the same as that of the linear harmonic oscillator (Note that the solution may be regular or singular depending on the form of $f(x)$).

A typical example is the perturbed Morse type oscillator 
\begin{eqnarray}
\label{cho40}
\ddot{x}+\lambda\dot{x}^2+\frac{\omega_0^2}{\lambda}(1-e^{-\lambda x})=0
\end{eqnarray}
corresponding to the Hamiltonian 
\begin{eqnarray}
\label{cho41}
H=\frac{1}{2\lambda^2}e^{-2\lambda x}p^2+\frac{1}{2}\omega_0^2(1-e^{-\lambda x}),
\end{eqnarray}
where the canonically conjugate momentum $p=\lambda^2e^{2\lambda x}\dot{x}$. The isochronos solution of (\ref{cho40}) is then given by
\begin{eqnarray}
\label{cho42}
x(t)=\frac{1}{\lambda}\ln [1-\lambda A\sin(\omega_0t+\delta)], \;\;0\leq A <\frac{1}{\lambda}.
\end{eqnarray}
The system can also be quantized as in the case of ML oscillator. One can find the general form of integrable systems of (\ref{cho04}) with three and two parameter symmetries whose integrals can be constructed from the symmetries. However, remarkably the ML oscillator possesses only one Lie point symmetry. Yet it is integrable, the reason being that it possesses the so called $\lambda$-symmetry which is essentially of nonlocal type\cite{Bhuneshwari:12}.

\section{Li\'enard Type II system: Modified Emden Equation and Generalizations - Nonlocal Transformations}
The standard Li\'enard system
\bea
\label{cho43}
\ddot{x}+h(x)\dot{x}+g(x)=0
\eea
can also be classified group theoretically\cite{Pandey:09}. It possesses Lie point symmetries which can be eight, three, two or one. The conditions for eight parameter symmetries to exist are 
\bea
\label{cho44}
h_{xx}=0,\qquad 3g_{xx}-2hh_x=0,
\eea
implying 
\bea
\label{cho45}
h(x)=ax+b,\qquad g(x)=\frac{2a}{3}(\frac{a}{6}x^3+\frac{b}{2}x^2+cx+d),
\eea
where $a, b, c$ and $d$ are constant parameters. In a similar way one can classify all the forms of (\ref{cho43}) possessing three and two point symmetries and their integrability can be established. The full equation (\ref{cho43}) for arbitrary form of $h(x)$ and $g(x)$ possesses at least one point symmetry whose integrability depends on the specific forms of $f(x)$ and $g(x)$.

\subsection{The Modified Emden Equation (MEE)}
An interesting case belonging to the family of Li\'enard type II equation possessing eight parameter Lie point symmetries satisfying the criteria (\ref{cho44}) or (\ref{cho45}) is the MEE\cite{mah:1985,Chandrasekar:05},
\bea
\label{cho46}
\ddot{x}+kx\dot{x}+\frac{k^2}{9}x^3+\omega_0^2x=0.
\eea
Eq. (\ref{cho46}) is linearizable under the nonlocal transformation
\bea
\label{cho46}
U(t)=x(t)e^{k\int xdt}
\eea
so that 
$\ddot{U}+\omega_0^2U=0$, $U=A\sin(\omega_0t+\delta)$, and 
\bea
\label{cho47}
\dot{x}-\frac{\dot{U}}{U}x+kx^2=0.
\eea
Consequently, solving the Riccati equation, we obtain the general solution as
\bea
\label{cho48}
x(t)=\frac{A\sin(\omega_0t+\delta)}{1-\frac{kA}{3\omega_0}\cos(\omega_0t+\delta)},\quad 0\le A<\frac{3\omega_0}{k}.
\eea
Note that the frequency is independent of the amplitude of oscillations or the solution (\ref{cho48}) is isochronous\cite{Calogero:08}. The integrability of system (\ref{cho46}) can be proved by finding two explicit time dependent integrals\cite{Chandrasekar:05}, from which one time independent integral can be identified. Consequently a nonstandard Hamiltonian and Lagrangian description can be developed for (\ref{cho46}) with the forms 
\begin{subequations}
\bea
\label{cho49}
&&L=\frac{27\lambda_1^3}{2k^2}
\bigg(\frac{1}{k\dot{x}+\frac{k^2}{3}x^2+3\lambda_1}\bigg)
+\frac{3\lambda_1}{2k}\dot{x}-\frac{9\lambda_1^2}{2k^2},\\
&&p=\left(\frac{-27\omega_0^6}{2k(k\dot{x}+\frac{k^2}{3}x^2+3\omega_0^2)^2}\right),\\
\label{cho50}
&&H=\frac{1}{2}F(p)x^2+U(p), \quad F(p)=\omega_0^2\left(1-\frac{2k}{3\omega_0^2}p\right),\\
&&U(p)=\frac{9\omega_0^4}{2k^2}\left(\sqrt{1-\frac{2k}{3\omega_0}p}-1\right)^2,
\eea
\end{subequations}
The isochronous time series and the phase space structure is shown in Fig. 2. Note that the trajectries are bounded in the upper half $(x,p)$ plane by the condition $p\leq\frac{3\omega_0^2}{2k}$, beyond which the trajectories become complex.

\begin{figure*}
\includegraphics[width=7.0cm,height=4.90cm]{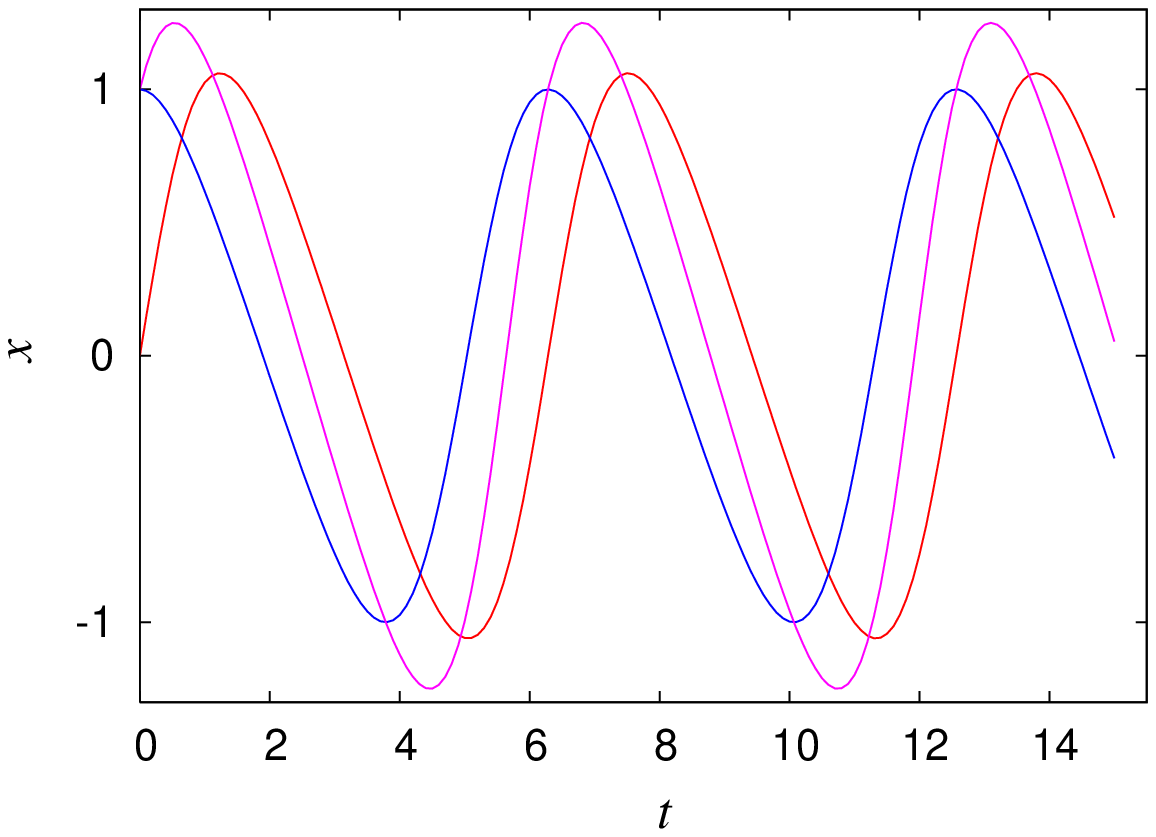}
\includegraphics[width=6.0cm,height=4.50cm]{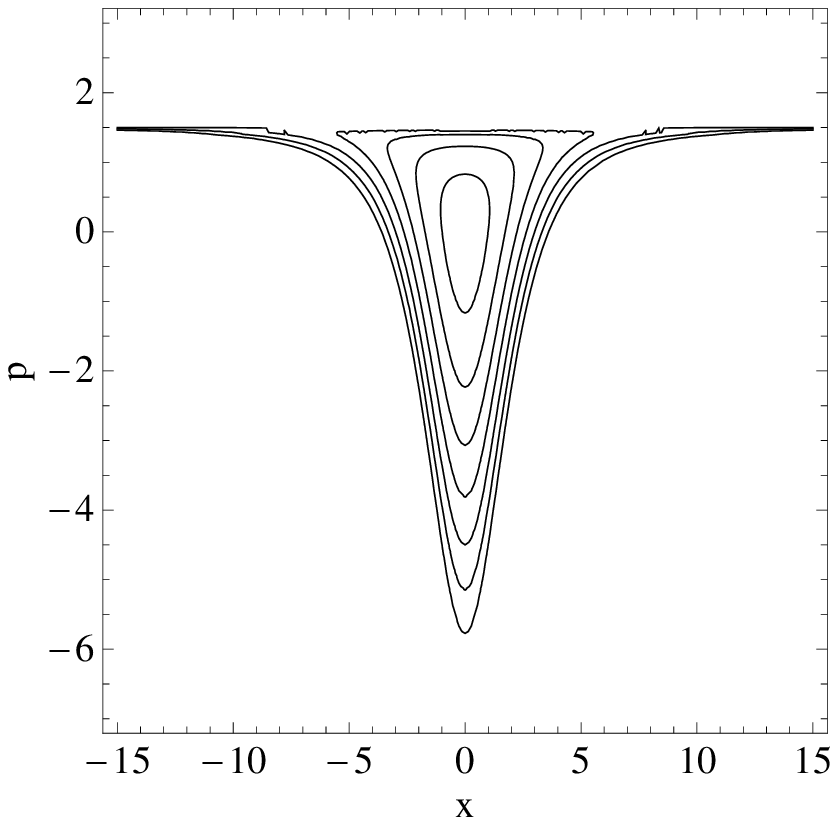}
\caption{(left) Time series and (right) $(x-p)$ phase plane structure of MEE.}
\label{fig2}
\end{figure*}

It is also of interest to compare the Hamiltonian structure of the Li\'enard type II MEE given by (\ref{cho50}) with the structure (\ref{cho13}) of the quadratic Li\'enard type I equation, where the role of $x$ and $p$ are interchanged. The quadratic structure in $x$ of the Hamiltonian (\ref{cho50}) allows one to quantize \cite{Chithika Ruby:12} the system exactly, now not in coordinate space but in momentum space. Before presenting the results briefly, we also note the fact that the MEE (\ref{cho46}) is invariant under the combined reversible transformation $x\rightarrow-x$ and $t\rightarrow -t$ so that (\ref{cho46}) and (\ref{cho50}) are {\it PT} symmetric systems but now for real dynamical variables. Note that the standard {\it PT} symmetric systems are {\it PT} invariant for complex dynamical variables\cite{Chithika Ruby:12,bender_r}. It appears that in this context system (\ref{cho46}) is unique.

To quantize the system (\ref{cho50}), one can rewrite the classical Hamiltonian in the position dependent mass form
\begin{eqnarray}
 H(x, p) = \frac{1}{2} \frac{x^2}{m(p)} + U(p),
\label{cho51}
\end{eqnarray}
where
\begin{eqnarray}
m(p) = \omega^{-2}\;\left(1 - \frac{2 k }{3 \omega^2} p\right)^{-1},
\quad U(p) = \frac{9\omega^4}{2 k^2} \left(\sqrt{1 - \frac{2 k}{3 \omega^2} p}-1\right)^2, \omega=\omega_0.\nonumber
\end{eqnarray}
We can now quantize the system in the momentum space using the so called von Roos ordering\cite{von} and with the operator replacement ${\displaystyle \hat{x} = i \hbar \frac{d}{dp}}$ in the time independent Schr\"{o}dinger equation,
\begin{equation}
\label{cho52}
H(\hat{x}, i \hbar \frac{d}{dp}) \Psi(x, p) = E \Psi(p),
\end{equation}
along with the boundary conditions $\Psi(-\infty) = \Psi\left(\frac{3 \omega^2}{2 k}\right) = 0$.
For admissible wavefunctions, one can show that the appropriate Hamiltonian for the present problem is 
\begin{eqnarray}
 H(\hat{x}, \hat{p}) = \frac{1}{2} \left[m^{-1/4}(\hat{p})\hat{x}m^{-1/2}(\hat{p})\hat{x}m^{-1/4}(\hat{p})\right] + U(\hat{p}),
\label{cho54}
\end{eqnarray}
so that the Schr\"{o}dinger equation becomes 
\begin{eqnarray}
 -\frac{\hbar^2 \omega^2 }{2}\left(1 - \frac{2 k}{3 \omega^2} p \right) \Phi'' - \frac{\hbar^2 k^2  }{ 24 \omega^2\left(1 - \frac{2 k}{3 \omega^2} p\right)} \Phi+ \frac{9 \omega^4}{2 k^2} \left(
\sqrt{1 - \frac{2 k}{3 \omega^2}p} - 1\right)^2 \Phi  = E \Phi,\nonumber
\label{cho55}
\end{eqnarray}
where $\left('= \frac{d}{d p}\right)$.
Note that the Hamiltonian (\ref{cho54}) is non-hermitian and nonsymmetric associated with its {\it PT} symmetric nature.
There are two sectors of solutions as given below.

(i) Case 1: The sector $-\infty < p \le \frac{3 \omega^2}{2 k}$:\\
\hspace{.1cm}
The {\it PT} invariant solution is
\begin{equation}
 \Phi_n(p) = \left\{\begin{array}{cc} \tilde{N}_n \left(1 - \frac{2k}{3\omega^2}p\right)^{1/4} \exp{\left(-\frac{9\omega^{3}}{2\;\hbar\;k^2}\left(1 - \frac{2k}{3\omega^2}p - 2\sqrt{1 - \frac{2k}{3\omega^2}p}\right)\right)}\nonumber \\
\hspace{3cm} \times H_{n}\left[\frac{3\omega^{3/2}}{\sqrt{\hbar}\;k}\left(\sqrt{1-\frac{2k}{3\omega^2}p} - 1\right)\right],
-\infty < p \le \frac{3 \omega^2}{2 k}, \label{sol2a} \\
                                  \hspace{-7cm} 0, \quad p \ge \frac{3 \omega^2}{2 k},
                                   \end{array}\right.
\end{equation}
with the associated eigenvalue spectrum as just that of the linear harmonic oscillator,
\begin{eqnarray}
 E_{n} &=& (n+\frac{1}{2})\;\hbar\; \omega, \qquad \quad n = 0, 1, 2, ....
\label{cho57}
\end{eqnarray}
In the above $H_n$'s are the Hermite polynomials and $N_n$ is the normalization constant.

(ii) Case 2: The sector $p>\frac{2\omega}{2k}$:
\begin{equation}
\Phi_n(p) = \left\{\begin{array}{cc} \tilde{\cal{N}}_n \left(\frac{2k}{3\omega^2}p - 1\right)^{1/4} \exp{\left(-\frac{9\omega^{3}}{2\;\hbar\;k^2}\left(\frac{2k}{3\omega^2}p -1 + i \;2\sqrt{\frac{2k}{3\omega^2}p -1}\right)\right)}\nonumber \\
\hspace{3cm} \times H_{n}\left[\frac{3\omega^{3/2}}{\sqrt{\hbar}\;k}\left(\sqrt{\frac{2k}{3\omega^2}p - 1} + i\right)\right],
              \quad p \ge \frac{3 \omega^2}{2 k}, \label{eig2a} \\
                                  \hspace{-7cm} \qquad\quad 0, \quad -\infty < p \le \frac{3 \omega^2}{2 k},
                                   \end{array}\right.,
\end{equation}
($\tilde{\cal{N}}_n$ is the normalization constant) with energy eigenvalues without a lower bound
\begin{equation}
\label{cho59}
E_n = - (n + \frac{1}{2}) \hbar \omega, \quad n = 0, 1, 2, 3, ... .
\end{equation}
Note that the eigenfunctions are no longer {\it PT} symmetric, even though the Hamiltonian (\ref{cho54})
 is, leading to a negative energy spectrum which is unbounded below, a property shared by other complex valued {\it PT} symmetric potentials\cite{bender_r}.

Thus one finds the above MEE and its quantized version turns out to be possessing unusual structures. Higher dimensional generalization of this system is indeed possible both at the classical and quantum levels, which will be presented elsewhere.

It is not only the specific form (\ref{cho46}) of MEE which is integrable. Even the generalized version\cite{feix:1997,Chandrasekar:07,Gladwin Pradeep:09a}
\bea
\label{cho60}
\ddot{x}+\alpha x\dot{x}+\beta x^3+\omega_0^2x=0,
\eea
where $\alpha$ and $\beta$ are arbitrary parameters, is completely integrable\cite{Chandrasekar:07} and time independent integrals of motion and solutions upto quadrature can be written down \cite{Chandrasekar:07}. However, in this case the structure of the integrals and solutions are more complicated. For details see Chandrasekar et al.\cite{Chandrasekar:07}. A simple way to look at the integrability of (\ref{cho60}) is that it is linearizable under the nonlocal transformation
\bea
y=\frac{x^2}{2}+\frac{\omega_0^2}{\lambda},\qquad d\tau=xdt
\label{cho61}
\eea
so that one obtains damped linear harmonic oscillator equation,
\bea
\label{cho62}
 \frac{d^2y}{d\tau^2}+\alpha\frac{dy}{d\tau}+\lambda y=0.
\eea
In fact one can generalize this result by considering a more general nonlocal transformation
\bea
y=\int f(x)dx,\qquad d\tau=\frac{f(x)}{g(x)}dt,
\label{cho63}
\eea
so that the general class of Li\'enard type equation III
\bea
\label{cho64}
\ddot{x}+\frac{g'(x)}{g(x)}\dot{x}^2+\alpha \frac{f(x)}{g(x)}\dot{x}+\lambda\frac{f(x)}{g(x)^2}\int f(x)dx=0
\eea
itself gets linearized to (\ref{cho62}). This allows one to associate a nonstandard Lagrangian and Hamiltonian description to (\ref{cho64}) as shown in Gladwin Pradeep et al.\cite{Gladwin Pradeep:09a}. For further discussion on nonstandard Lagrangian/Hamiltonian formulation, see for example 
refs.\cite{Musielak:08,Cie:10}

\section{Higher dimensional coupled integrable versions of MEE}
An interesting two dimensional generalization of MEE given by (\ref{cho46}) can be identified\cite{Gladwin Pradeep:09b} as 
\begin{eqnarray}
&&\ddot{x}=-2(k_1x+k_2y)\dot{x}-(k_1\dot{x}+k_2\dot{y})x-
(k_1x+k_2y)^2x-\lambda_1 x,\nonumber\\
&&\ddot{y}=-2(k_1x+k_2y)\dot{y}-(k_1\dot{x}+k_2\dot{y})y-
(k_1x+k_2y)^2y-\lambda_2 y.
\label{cho65}
\end{eqnarray}
Eq. (\ref{cho65}) can be linearized under the nonlocal transformation
\begin{eqnarray}
\label{cho66}
 U=xe^{\int(k_1x+k_2y)dt}, \;\;V=ye^{\int(k_1x+k_2y)dt}.
\eea
so that
\begin{eqnarray}
\label{cho67}
\ddot{U}+\lambda_1U=0,\;\;\ddot{V}+\lambda_2V=0.
\eea
From (\ref{cho66}), one can also identify a set of coupled Riccati equations,
\bea
\label{cho68}
\dot{x}=\frac{\dot{U}}{U}x-k_1x^2-k_2xy,\quad
\dot{y}=\frac{\dot{V}}{V}y-k_1xy-k_2y^2.
\end{eqnarray}
Solving the system (\ref{cho68}) one can obtain the explicit oscillatory solutions,
\begin{eqnarray}
\label{cho69}
&&x(t)=\frac{A\sin(\omega_1 t+\delta_1)}
{1-\frac{Ak_1}{\omega_1}\cos(\omega_1 t+\delta_1)
-\frac{Bk_2}{\omega_2}\cos(\omega_2 t+\delta_2)},\nonumber\\
&&y(t)=\frac{B\sin(\omega_2 t+\delta_2)}
{1-\frac{Ak_1}{\omega_1}\cos(\omega_1 t+\delta_1)
-\frac{Bk_2}{\omega_2}\cos(\omega_2 t+\delta_2)},\,\,\nonumber\\
&&\hspace{7cm}\bigg|\frac{Ak_1}{\omega_1}+\frac{Bk_2}{\omega_2}\bigg|<1,\nonumber
\end{eqnarray}
where $\omega_j=\sqrt{\lambda_j},\;j=1,2,\,\bigg|\frac{Ak_1}{\omega_1}+\frac{Bk_2}{\omega_2}\bigg|<1$. 
Note that the solution may be periodic or quasiperiodic depending on the value of the ratio $\omega_1/\omega_2$, that is whether it is rational or irrational. Typical solutions are plotted in Fig. 3.

\begin{figure*}
\includegraphics[width=12.50cm,height=7.90cm]{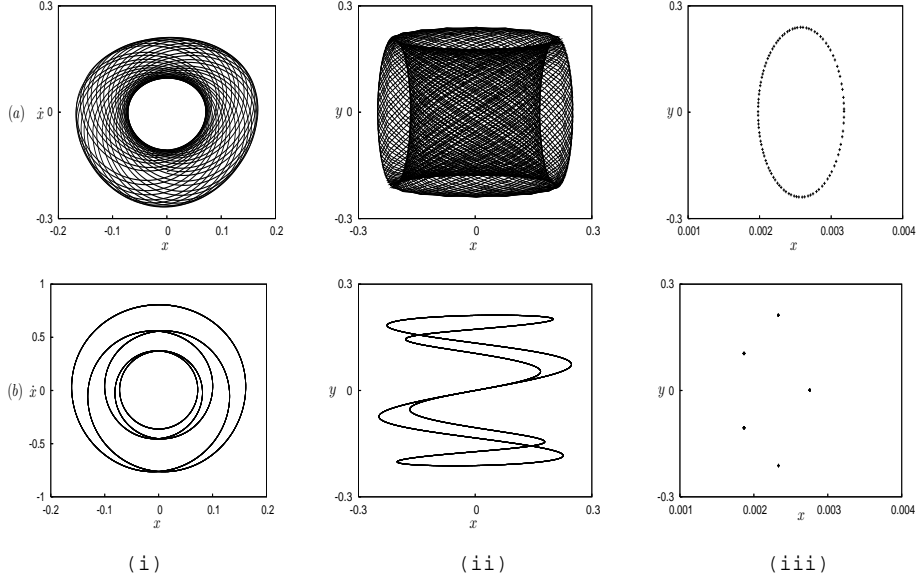}
\caption{(a) Quasi-periodic oscillations ($\omega_1=1$ and $\omega_2=\sqrt{2}$) (b) periodic oscillations ($\omega_1=1$ and $\omega_2=5$) of Eq. (\ref{cho65}). Here 
(i) phase space plot (ii) Configuration space plot and (iii) Poincar\'e SOS.}
\label{fig3}
\end{figure*}

The system (\ref{cho60}) also admits four independent integrals, two of which are time independent and the remaining two are time dependent:
\begin{subequations}
\begin{eqnarray}
&&I_1=\frac{(\dot{x}+(k_1x+k_2 y)x)^2+\lambda_1 x^2}
{\left[\frac{k_1}{\lambda_1}(\dot{x}+(k_1x+k_2y)x)+\frac{k_2}{\lambda_2}(\dot{y}+(k_1x+k_2y)y)+1\right]^2},\label {cho70a}\\
&&
I_2=\frac{(\dot{y}+(k_1x+k_2 y)y)^2+\lambda_2 y^2}
{\left[\frac{k_1}{\lambda_1}(\dot{x}+(k_1x+k_2y)x)+\frac{k_2}{\lambda_2}(\dot{y}+(k_1x+k_2y)y)+1\right]^2},\label {cho70b}
\end{eqnarray}
\label {cho70}
\end{subequations}
\begin{eqnarray}
&&I_3=\left\{
\begin{array}{ll}
\tan^{-1}\left[\frac{\sqrt{\lambda_1}x}{\dot{x}+(k_1x+k_2y)x}\right]
-\sqrt{\lambda_1}\,t,&\lambda_1>0
\\
\frac{e^{2\sqrt{|\lambda_1|}t}(\dot{x}+(k_1x+k_2y)x-\sqrt{|\lambda_1|}x)}
{\dot{x}+(k_1x+k_2y)x+\sqrt{|\lambda_1|}x},&\lambda_1<0,
\end{array}
\right.\label {cho71}
\end{eqnarray}
and
\begin{eqnarray}
&&I_4=\left\{
\begin{array}{ll}
\tan^{-1}\left[\frac{\sqrt{\lambda_2}y}{\dot{y}+(k_1x+k_2y)y}\right]
-\sqrt{\lambda_2}\,t,&\lambda_2>0
\\
\frac{e^{2\sqrt{|\lambda_2|}t}(\dot{y}+(k_1x+k_2y)x-\sqrt{|\lambda_2|}y)}
{\dot{y}+(k_1x+k_2y)y+\sqrt{|\lambda_2|}y},&\lambda_2<0.
\end{array}
\right.\label {cho72}
\end{eqnarray}
The system (\ref{cho60}) does admit a singular Lagrangian for the case 
$\lambda_1=\lambda_2=\lambda$ as
\bea
\label {cho73}
L=\frac{1}{\left[k_1(\dot{x}+(k_1x+k_2y)x)+k_2(\dot{y}+(k_1x+k_2y)y)+\lambda\right]}.
\eea
The problem of constructing appropriate Lagrangian and Hamiltonian to (\ref{cho65}) still remains an open problem.
One can generalize the above results to a system of N-coupled MEEs,
\begin{eqnarray}
\label {cho74}
\ddot{x}_i+ 2\sum_{j=1}^Nk_jx_j\dot{x}_i+\sum_{j=1}^Nk_j\dot{x}_jx_i
+(\sum_{j=1}^Nk_jx_j)^2x_i+\lambda_ix_i=0,\;i=1,2,...,N,
\end{eqnarray}
and obtain the explicit periodic solutions. The associated 2N integrals turn out to be 
\begin{eqnarray}
&&
I_{1i}=\frac{(\dot{x}_i+\sum_{j=1}^N(k_jx_j)x_i)^2+\lambda_ix_i^2}
{\left[\sum_{j=1}^N\left[\frac{k_j}{\lambda_j}(\dot{x}_j+\sum_{n=1}^N(k_nx_n)x_j)\right]+1\right]^2},\label{cho75}\\
&&
I_{2i}=\tan^{-1}\left[\frac{\sqrt{\lambda_i}x_i}
{\dot{x}_i+\sum_{j=1}^N(k_jx_j)x_i}\right]-\sqrt{\lambda_i} t,\,\,\,\lambda_i>0,\label{cho76}
\end{eqnarray}
We also note here that one can make appropriate contact-type transformations to (\ref{cho74}) which maps them onto a system of N uncoupled harmonic oscillators.
\subsection{Generalized nonlocal transformations and integrable N-coupled dynamical systems}
Consider the set of uncoupled linear harmonic oscillators
\bea
\label{cho77}
\ddot{U}_i+\omega_i^2U_i=0,\;\;i=1,2,\ldots,N.
\eea
Under the nonlocal transformation
\bea
\label{cho78}
U_i=x_ie^{\int f_i(x_1,x_2,\ldots,x_N)dt},
\eea
Eqs. (\ref{cho77}) gets transformed to\cite{Chandrasekar:12}
\bea
\label{cho79}
\ddot{x}_i+\sum_{j=1}^N f_i^{(j)}(\overline{x})x_i\dot{x}_j+f_i(\overline{x})\dot{x}_i+f_i^2(\overline{x})x_i+\omega_i^2x_i=0,\,\,i=1,2,\ldots,N,
\eea
where $f_i^{(j)}=\frac{df_i}{dx_j}$ and $\overline{x}=(x_1,x_2,\ldots, x_N)$. Then using the Riccati connection, we can obtain the set of first order ODEs,
\bea
\label{cho80}
\dot{x}_i=\frac{\omega_i}{\tan(\omega_it+\delta_i)}x_i-f_i(x_1,x_2,\ldots,x_N)x_i,\quad i=1,2,\ldots N.
\eea
($\delta_i$'s : $N$-integration constants). For the special choice
\begin{eqnarray}
&&f_k=f_N(\bar{x})=f(\bar{x})=f(x_1, x_2, \ldots x_N), \; k=1,2,\ldots, N-1, \label{cho81}
\eea
$(N-1)$ integrals can be obtained from the relations 
\bea
\frac{x_k}{x_N}=I_{k}\frac{\sin(\omega_kt+\delta_k)}{\sin(\omega_Nt+\delta_N)}=h_k(t), \;\;k=1,2,\ldots, N-1
\label{cho82}
\end{eqnarray}
where $I_k$'s are the (N-1) independent integrals.

Then the problem (\ref{cho79}) reduces to the problem of solving a single first order ODE\cite{Chandrasekar:12}: 
\begin{eqnarray}
\label{cho83}
 \dot{x}_N=\frac{\omega_N x_N}{\tan(\omega_Nt+\delta_N)}-f(\bar{h}(t),x_N)x_N.
\end{eqnarray}
The associated integrals can be explicitly given as
\begin{eqnarray}
\label{cho84}
I_k^2&=&\frac{(\omega_k^2x_k^2+(\dot{x}_k+f(\bar{x})x_k)^2)}{(\omega_N^2x_N^2+
(\dot{x}_N+f(\bar{x})x_N)^2)}
, \;\;k=1,2,\ldots, N-1.
\end{eqnarray}
The remaining integral $I_N$ can be obtained by solving the Riccati equation (\ref{cho83}) for appropriate forms of $f$.
One can extend this procedure to analyse even nearby nonintegrable 
systems\cite{Chandrasekar:12}.
\section{Singular and nonsingular isochronous Hamiltonian systems}
In the following we briefly discuss the procedure to deduce singular\cite{singular2} and nonsingular Hamiltonian 
systems associated with a class of isochronous systems\cite{Calogero:07,Calogero:08a}.
\subsection{Singular isochronous Hamiltonian systems}
Let us consider an $N$-dimensional system with a Hamiltonian of the form\cite{Gladwin Pradeep:12c} involving velocity dependent potentials,
\begin{eqnarray}
\tilde H(\underline {p},\underline {q};\Omega)
&=&\frac{1}{2}\left[\left\{\sum_{n=1}^{N}a_{n}p_{n}\left(\frac{\partial Q(\underline {q})}{\partial q_{n}}\right)^{-1}\right\}^2+\Omega^2Q(\underline{q})^2\right],
\label{isoham1}
\end{eqnarray}
where $a_n$ and $\Omega$ are constants, $\underline{p}=(p_{1},p_{2},...,p_{N})$ and $\underline {q}=(q_{1},q_{2},...,q_{N})$ and $Q(\underline {q})$ is an arbitrary function of the canonical coordinates $q'_{n}s,\,\, n=1,2,...N$.  Hamiltonian (\ref{isoham1}) results in the following $2N$ coupled first order canonical equations of motion for the canonical coordinates $q_n$ and $p_n$,
\begin{subequations}
\label{isoeq1-can}
\begin{eqnarray}
&&\dot{q}_n=a_n\left[\frac{\partial Q(\underline{q})}{\partial q_n}\right]^{-1}H,\qquad n=1,2,\ldots,N\label{isoeq1-can1}\\
&&\dot{p}_n=H\sum_{m=1}^N\left\{a_mp_m\left[\frac{\partial Q(\underline{q})}{\partial q_m}\right]^{-2}\left[\frac{\partial^2Q(\underline{q})}{\partial q_m\partial q_n}\right]\right\}-\Omega^2Q(q)\frac{\partial Q(\underline{q})}{\partial q_n}.\label{isoeq1-can2}
\end{eqnarray}
\end{subequations}
The corresponding  Newton's equation of motion is obtained by differentiating (\ref{isoeq1-can1}) with respect to $t$ and using (\ref{isoeq1-can2}) for $\dot{p}_n$. It is given in the form
\begin{eqnarray}
\label{isoeq1}
\frac{\partial Q(\underline{q})}{\partial q_{n}}\ddot q_{n}+\sum_{m=1}^{N}\bigg(\dot q_{n}\dot q_{m}\frac{\partial^{2}Q(\underline{q})}{\partial q_{n}q_{m}}\bigg)+\Omega^{2} a_{n}Q(\underline{q})=0,n=1,2,\ldots,N.
\end{eqnarray}
However, one can easily check that not all the coordinates $q_i,\,\,i=1,2,\ldots,N$ are independent: there are $(N-1)$ holonomic constraints existing between them\cite{Gladwin Pradeep:12c}:
\begin{eqnarray}
\int dq_1\frac{\partial Q}{\partial q_1}-\int dq_j\frac{\partial Q}{\partial q_j}=C_j,\quad j=2,3,\ldots,N,\label{int-constraint}
\end{eqnarray}
where $C_j$'s are constants.  Equation (\ref{int-constraint}) obviously constitutes a set of $(N-1)$ holonomic constraints on the coordinates $q_i$.  One can easily check that the Hamiltonian (\ref{isoham1}) is indeed {\it singular}: The Hessian  $\displaystyle\vline\frac{\partial^2 \tilde{H}}{\partial p_i\partial p_j}\vline=0$ and is of rank one only. 

Interestingly the Newton's equation of motion (\ref{isoeq1}) admits a \emph{nonsingular} Hamiltonian too:
\begin{eqnarray}
\tilde{H}=\frac{1}{2}\left(\sum_{i=1}^Na_i^2p_i^2\left[\frac{\partial Q(\underline{q})}
{\partial q_i}\right]^{-2}+\Omega^2Q(\underline{q})^2\right),
\end{eqnarray}
with the associated canonical equations,
\begin{subequations}
\begin{eqnarray}
\dot{q}_i=a_i^2p_i\left[\frac{\partial Q(\underline{q})}{\partial q_i}\right]^{-2},\;\;
\dot{p}_i=\sum_{j=1}^Na_j^2p_j^2\left[\frac{\partial Q(\underline{q})}{\partial q_j}\right]^{-3}\frac{\partial^2 Q(\underline{q})}{\partial q_j \partial q_i}-\Omega^2Q(\underline{q})\frac{\partial Q}{\partial q_i}.
\end{eqnarray}
\end{subequations}

To be more specific, we consider the choice
\begin{eqnarray}
Q(\underline{q})=\sum_{m=1}^N b_mq_m^{k_m},\label{calogero-form}
\end{eqnarray}
where $b_m$'s are arbitrary real parameters and $k_m$'s are such that $(1/k_m)$'s are positive integers.
For the Newton's equation (\ref{isoeq1}) \emph{without} the constraints (\ref{int-constraint}) one obtains the general solution as
\begin{eqnarray}
q_n(t)=q_n(0)\left(1+\frac{a_n}{b_n}\frac{1}{(q_n(0))^{k_n}}\left(\frac{H(0)}{\Omega}\sin\Omega t+Q(0)(\cos\Omega t-1)\right)+C_n(0)t\right)^{\frac{1}{k_n}},\label{calogero-gen}
\end{eqnarray}
where $q_n(0)'s$, $\,H(0)$, and $C_n(0)$, $\sum_{n=1}^NC_n=0$, $n=1,2,\ldots,N$ are integration constants fixed by the initial condition and $Q(0)=\sum_{m=1}^Nb_mq_m(0)^{k_m}$. But the solution (\ref{calogero-gen}) is unbounded and nonisochronic. On the other hand, subject to the $(N-1)$ constraints (\ref{int-constraint}), the Newton's equation admits the $(N+1)$ parameter bounded, isochronous solution
\begin{eqnarray}
&&\hspace{-1cm}q_{n}(t)=q_{n}(0)\bigg(1+\frac{a_{n}}{b_{n}}\frac{1}{(q_{n}(0))^{k_{n}}}\bigg[H(0)\frac{\sin(\Omega t)}{\Omega}+Q(0)(\cos(\Omega t)-1)\bigg]\bigg)^\frac{1}{k_{n}}, \label{isosol1}
\end{eqnarray}
which is also the solution of the Hamilton's equations (\ref{isoeq1-can}).  The solution (\ref{isosol1}) is isochronous and bounded but corresponds to the singular Hamiltonian of the form (\ref{isoham1}).

Our analysis clearly shows that for the singular Hamiltonian systems (\ref{isoham1}), the equivalent Newton's equation is a holonomic constrained system (with $(N-1)$ constraint conditions) admitting isochronous oscillatory solution as the general solution.  Consequently, the associated system possesses only one independent coordinate variable. However, the nonsingular Hamiltonian system admits only unbounded solution as the general solution.   

In the following, we describe a procedure to modify this system such that the new system with $N$-degrees of freedom admits isochronous oscillations and nonsingular Hamiltonian.

\subsection{Systematic method to construct higher dimensional isochronous systems}
Let us define the modified Hamiltonian for a two-dimensional system as
\begin{eqnarray}
\tilde{H}
&=&\frac{1}{2}\bigg[\frac{\big(p_{1} Q_{2q_{2}}-p_{2} Q_{2q_{1}}\big)^2}{\Delta^2}+\frac{\big(p_{2} Q_{1q_{1}}-p_{1} Q_{1q_{2}}\big)^2}{\Delta^2}\nonumber\\&&\qquad\qquad\qquad\qquad\qquad
+\Omega_{1}^2Q_{1}(q_{1},q_{2})^2+\Omega_{2}^2Q_{2}(q_{1},q_{2})^2\bigg]
\label{isoham2}
\end{eqnarray}
From the Hamilton's equations of motion we get the following system of constraint free two coupled second order ODEs,
\begin{eqnarray}
\ddot{q}_i=\frac{1}{\Delta}\left(\sum_{k=1}^2\sum_{j=1}^2A_{ijk} \dot{q}_j\dot{q}_k+B_i\right),\quad i=1,2,\label{two-coupled-eq}
\end{eqnarray}
where
\begin{eqnarray}
A_{ijk}=(-1)^{i+1}\left|\begin{array}{ccc}
Q_{1q_{i+1}}&Q_{2q_{i+1}}\\
Q_{1q_{j}q_k}&Q_{2q_{j}q_k}
\end{array}\right|,
\quad
B_{i}=(-1)^i\left|\begin{array}{cccc}
Q_{1}\Omega_1^2&Q_{2}\Omega_2^2\\
Q_{1q_{i+1}}&Q_{2q_{i+1}}
\end{array}\right|,2+i=i.
\end{eqnarray}
In order to obtain the explicit general solution of (\ref{two-coupled-eq}) one has to fix the form of $Q_1$ and $Q_2$ in the above equation such that the resultant solutions are analytic and single valued.  For example, we can make the choice
\begin{eqnarray}
Q_{1}=k_{1}q_{1}^{r_{1}}+k_{2}q_{2}^{r_{2}}\quad Q_{2}=k_{3}q_{1}^{r_{1}}+k_{4}q_{2}^{r_{2}},\label{harmq1}
\end{eqnarray}
where $r_1$ and $r_2$ are such that $(1/r_1)$ and $(1/r_2)$ are positive integers so that the resultant solution is single valued and analytic. 

The general solution of the system (\ref{two-coupled-eq}) can be obtained as
\begin{subequations}
\label{isosol2}
\begin{eqnarray}
&&q_1=\left(\frac{\frac{Bk_{2}}{\Omega_2}\sin(\Omega_2t+\delta_2)-\frac{Ak_{4}}{\Omega_1}\sin(\Omega_1t+\delta_1)}{k_{2}k_{3}-k_{1}k_{4}}\right)^{\frac{1}{r_{1}}}, (k_2k_3-k_1k_4)\ne 0,  \\
&& q_2=\left(\frac{\frac{Ak_{3}}{\Omega_1}\sin(\Omega_1t+\delta_1)-\frac{Bk_{1}}{\Omega_2}\sin(\Omega_2t+\delta_2)}{k_{2}k_{3}-k_{1}k_{4}}\right)^{\frac{1}{r_{2}}}.   
\end{eqnarray}
\end{subequations}
The obtained solution (\ref{isosol2}) is analytic and bounded and exhibits oscillatory behaviour for the choice $1/r_1$ and $1/r_2$ which are positive integers.

One can generalize the procedure of constructing isochronous Hamiltonian systems to $N$ degrees of freedom system. We find the following system of $N$ coupled second order ODEs,
\begin{eqnarray}
\ddot{q}_i=\frac{-1}{\Delta}\left(\sum_{k=1}^N\sum_{j=1}^N A_{ijk}\dot{q}_j\dot{q}_k+B_{i}\right),\quad i=1,2,\ldots,N,\,\,N>2\label{ndimsys}
\end{eqnarray}
where $A_{ijk}$ and $B_{ij}$, $j,k=1,2,\ldots,N$ are determinants of the form
\begin{eqnarray}
A_{ijk}=\left|\begin{array}{cccc}
Q_{1q_{i+1}}&Q_{2q_{i+1}}&\ldots&Q_{Nq_{i+1}}\\
Q_{1q_{i+2}}&Q_{2q_{i+2}}&\ldots&Q_{Nq_{i+2}}\\
\vdots&\ddots&\ddots&\vdots\\
Q_{1q_{N}}&Q_{2q_{N}}&\ldots&Q_{Nq_{N}}\\
Q_{1q_{1}}&Q_{2q_{1}}&\ldots&Q_{Nq_{1}}\\
\vdots&\ddots&\ddots&\vdots\\
Q_{1q_{i-1}}&Q_{2q_{i-1}}&\ldots&Q_{Nq_{i-1}}\\
Q_{1q_{j}q_{k}}&Q_{2q_{j}q_{k}}&\ldots&Q_{Nq_{j}q_{k}}
\end{array}\right|_{N\times N},
B_{i}=\left|\begin{array}{cccc}
Q_1\Omega_1^2&Q_2\Omega_2^2&\ldots&Q_{N}\Omega_N^2\\
Q_{1q_{i+1}}&Q_{2q_{i+1}}&\ldots&Q_{Nq_{i+1}}\\
Q_{1q_{i+2}}&Q_{2q_{i+2}}&\ldots&Q_{Nq_{i+2}}\\
\vdots&\ddots&\ddots&\vdots\\
Q_{1q_{N}}&Q_{2q_{N}}&\ldots&Q_{Nq_{N}}\\
Q_{1q_{i-1}}&Q_{2q_{i-1}}&\ldots&Q_{Nq_{i-1}}
\end{array}\right|_{N\times N},\nonumber
\end{eqnarray} 
admit isochronous oscillations for appropriate choices of the determinants $A$ and $B$. For the general solution of (\ref{ndimsys}), canonical variables $q_{i},\,i=1,2,...,N$, evolve periodically with a fixed period $T$ when $\Omega_i$'s are commensurate, for appropriate forms of $Q_{i}(q_{1},q_{2},\ldots,q_{N})$ such that the resultant solutions $q_i$'s are analytic and single valued\cite{Gladwin Pradeep:12c}.

\section{Generalized Li\'enard type  $III$ and IV equations}
Let us consider a damped linear harmonic oscillator,
\begin{eqnarray}            
\ddot{U}+c_1 \dot{U}+c_2 U=0,
\label {nld01}
\end{eqnarray}
where $c_1$ and $c_2$ are arbitrary parameters. Now we consider a nonlocal 
transformation of the form
\begin{eqnarray}            
U=x^ne^{\int_0^t{(\beta(t') x^m+\gamma(t'))}dt'},
\label {nld02}
\end{eqnarray}
where $n$ and $m$ are constants and $\beta(t)$ and $\gamma(t)$ are 
arbitrary functions of $t$, 
and substitute it into (\ref{nld01}) so that the latter becomes a nonlinear 
second order ODE of the form\cite{Chandrasekar:06}
\begin{eqnarray}            
\ddot{x}+(n-1) \frac{\dot{x}^2}{x}+\frac{\beta^2}{n} x^{2m+1}+b_1(t,x)\dot{x}
+b_2(t)x^{m+1}+b_3(t)x=0,
\label {nld03}
\end{eqnarray}
where 
\begin{eqnarray}   
&&b_1(t,x)=\frac{1}{n}\bigg(2n\gamma+nc_1
+(m+2n)\beta x^m\bigg),\nonumber\\
&&b_2(t)=\frac{1}{n}\bigg( \dot{\beta}+2\gamma \beta
+c_1\beta \bigg),\;\;
b_3(t)=\frac{1}{n}\bigg(\dot{\gamma}+\gamma^2+\gamma c_1
+c_2 \bigg).
\label {nld04}
\end{eqnarray}
 From equation~(\ref{nld02}) we get the first order ODE,
\begin{eqnarray}            
\dot{x}=\bigg(\frac{\dot{U}}{U}-\gamma(t)\bigg)\frac{x}{n}
-\frac{\beta(t)}{n}x^{m+1}.
\label {nld06}
\end{eqnarray}
Solving equation (\ref{nld06})  we get the general solution for Eq. (\ref{nld03}) in the form
\begin{eqnarray}            
 x(t)=
e^{\frac{1}{n}\int_0^t(\hat{a}(t')-\gamma(t'))dt'}
\bigg[C+\frac{m}{n}\int_0^t\bigg(\beta(t')e^{\frac{m}{n}\int_0^{t'}
(\hat{a}(t'')-\gamma(t''))dt''}\bigg)dt'\bigg]^{\frac{-1}{m}} .
\label {nld07}
\end{eqnarray}

Interestingly,  one can introduce the independent variable $t$ using the general nonlocal transformation (\ref{cho63}), that is, $y=\int f(x,t)dx,\qquad d\tau=\frac{f(x,t)}{g(x,t)}dt$. Then the Li\'enard type III equation becomes the general class of Li\'enard type IV equation of the form
\bea
\label{cho64a}
\ddot{x}+\frac{g_x}{g}\dot{x}^2+\alpha \frac{f+g_t}{g}\dot{x}+\lambda\frac{f}{g^2}\int fdx=0.
\eea
Special cases of the celebrated Gambier equation\cite{GR,GRL,Guga1} can be related to the above system. The most general form of second-order Gambier equation\cite{Gambier:10,GR,GRL,Guga1} is described by the following form
\begin{eqnarray}
\label{pain2}
w''={r-1\over r}{w'^2\over w}+a{r+2\over r}ww' +bw'-{r-2\over r}{w'\over w}\sigma-{a^2 \over r}x^3 \nonumber\\
\phantom{\ddot{x}=}{} +
(a'-ab)x^2
+ \Big(cr-{2a\sigma\over r}\Bigg)w-b\sigma-{\sigma^2\over rw},
\end{eqnarray}
where $a$, $b$ and $c$ are functions of the independent variable $z$, $r$ is an integer and $\sigma$ is a constant. Gambier equation describes all the linearisable equations (not necessarily under point transformations but involves more general transformations) of the Painlev\'e-Gambier list 
by making appropriate limits in their coefficients \cite{Gambier:10,GRL,Guga1}. The above results can be generalized to higher order ODEs also. For details see ref.\cite{Gladwin Pradeep:10}.

Next, let us consider a set of two uncoupled damped linear harmonic oscillators
\begin{eqnarray}
\label {nld08}
\ddot{U}+c_{11}\dot{U}+c_{12}U=0,\;\;\;\;\;
\ddot{V}+c_{21}\dot{V}+c_{22}V=0.
\end{eqnarray}
Introducing a nonlocal transformation,
\begin{eqnarray}
\label {nld09}
U=x^{\alpha}e^{\int f(x,y,t)dt}, \quad V=y^{\beta}e^{\int g(x,y,t)dt},
\end{eqnarray}
where $f$ and $g$ are two arbitrary functions of their arguments, in (\ref{nld08}) we obtain a set of two coupled second order nonlinear ODEs of the form\cite{Gladwin Pradeep:10}
\begin{eqnarray}
\label {nld10}
&& \ddot{x}+(\alpha-1)\frac{\dot{x}^2}{x}+(2f+c_{11})\dot{x}
+\frac{x}{\alpha}\left[f^2+c_{11}f+c_{12}+\dot{f}\right]=0,\,\nonumber\\
&& \ddot{y}+(\beta-1)\frac{\dot{y}^2}{y}+(2g+c_{21})\dot{y}+\frac{y}{\beta}\left[g^2+c_{21}g+c_{22}+\dot{g}\right]=0,
\end{eqnarray}
The solution of equation (\ref{nld10}) can be obtained from the solution of a system of two
first order coupled nonlinear, nonautonomous ODEs of the form
\begin{eqnarray}
\dot{x}=\frac{x}{\alpha}\left[\frac{\dot{U}}{U}-f(x,y,t)\right],
\qquad\dot{y}=\frac{y}{\beta}\left[\frac{\dot{V}}{V}-g(x,y,t)\right].
\end{eqnarray}

This analysis can be generalized in principle to a system of arbitrary 2-coupled $l$th order ODEs and classes of solvable ones from the linear ODEs can be identified. Let us consider a system of two uncoupled linear ODEs
\begin{eqnarray}
\bigg(\frac{d^l}{dt^l}+c_{11}\frac{d^{(l-1)}}{dt^{(l-1)}}
+\ldots+c_{1l-1}\frac{d}{dt}+c_{1l}\bigg)U(t)=0,\nonumber\\
\bigg(\frac{d^l}{dt^l}+c_{21}\frac{d^{(l-1)}}{dt^{(l-1)}}
+\ldots+c_{2l-1}\frac{d}{dt}+c_{2l}\bigg)V(t)=0,\label {nld11}
\end{eqnarray}
where $c_{ij}$’s, $i = 1,2,$ $j = 1,2,...l,$ are arbitrary constants. The nonlocal transformation (\ref{nld09}) connects (\ref{nld11}) to the set of coupled nonlinear ODEs of the form
\begin{eqnarray}
\bigg(D_{1}^{(l)}+c_{11}D_{1}^{(l-1)}+\ldots+c_{1l-1}D_{1}^{(1)}
+c_{1l}\bigg)x=0,\nonumber\\
\bigg(D_{2}^{(l)}+c_{21}D_{2}^{(l-1)}+\ldots+c_{2l-1}D_{2}^{(1)}
+c_{2l}\bigg)y=0,\label {nld12}
\label {nlv02}
\end{eqnarray}
where $D_{1}^{(l)} =(\frac{d}{d t}+f(x,y,t))^l$ and
$D_{2}^{(l)} =(\frac{d}{d t}+g(x,y,t))^l$. The solution of Eq. (\ref{nld12}) can be deduced from the nonlocal transformation (\ref{nld09}) and the solution of the system of linear ODEs (\ref{nld11}) only for specific forms of f (x, y, t) and g(x, y, t)\cite{ref10}. These results can be further generalized to $n$th order systems as well.

Using the nonlocal connection between linear and nonlinear ODEs one can generate the following integrable chains of ODEs:\\
(a)  Coupled Ricatti chain:
\begin{eqnarray}
&&\mathbb{D}_{R}^0
\left(
\begin{array}{l}
x_1\\
x_2
\end{array}
\right)\,\,\Rightarrow
\begin{array}{l}
x_1=0,\\
x_2=0.
\end{array}\label{chaineq1}\\
&&\mathbb{D}_{R}^1
\left(
\begin{array}{l}
x_1\\
x_2
\end{array}
\right)
\,\,\Rightarrow
\begin{array}{l}\dot{x}_1+(a_1x_1+a_2x_2)x_1=0,\\
\dot{x}_2+(a_1x_1+a_2x_2)x_2=0.
\end{array}\\
&&\mathbb{D}_{R}^2
\left(
\begin{array}{l}
x_1\\
x_2
\end{array}
\right)\,\,\Rightarrow
\begin{array}{l}
\ddot{x}_1+2(a_1x_1+a_2x_2)\dot{x}_1+(a_1\dot{x}_1+a_2\dot{x}_2)x_1
+(a_1x_1+a_2x_2)^2x_1=0,\\
\ddot{x}_2+2(a_1x_1+a_2x_2)\dot{x}_2+(a_1\dot{x}_1+a_2\dot{x}_2)x_2
+(a_1x_1+a_2x_2)^2x_2=0.\\
\end{array}\label{cmee}\\
&&\mathbb{D}_{R}^3
\left(
\begin{array}{l}
x_1\\
x_2
\end{array}
\right)\Rightarrow
\begin{array}{l}
\dddot{x}_1+(3(a_1x_1+a_2x_2))\ddot{x}_1+x(a_1\ddot{x}_1+a_2\ddot{x}_2)
+3(a_1\dot{x}_1+a_2\dot{x}_2)\dot{x}_1\\
\quad
+(a_1x_1+a_2x_2)(3\dot{x}_1(a_1x_1+a_2x_2)
+3x_1(a_1\dot{x}_1+a_2\dot{x}_2)
\\
\quad+(a_1x_1+a_2x_2)^2x_1)=0,\\
\dddot{x}_2+(3(a_1x_1+a_2x_2))\ddot{x}_2
+y(a_1\ddot{x}_1+a_2\ddot{x}_2)
+3(a_1\dot{x}_1+a_2\dot{x}_2)\dot{x}_2
\\
\quad
+(a_1x_1+a_2x_2)(3\dot{x}_2(a_1x_1+a_2x_2)
+3x_2(a_1\dot{x}_1+a_2\dot{x}_2)
\\
\quad+(a_1x_1+a_2x_2)^2x_2)=0,
\end{array}\label{chazy}
\end{eqnarray}
and so on, where $\mathbb{D}_{R}^l=\left(\frac{d}{dt}+(a_1x_1+a_2x_2)\right)^l$.

(b) Coupled Abel chain:
\begin{eqnarray}
&&\mathbb{D}_{A}^0
\left(
\begin{array}{l}
x_1\\
x_2
\end{array}
\right)\,\,\Rightarrow
\begin{array}{l}
x_1=0,\\
x_2=0.\label{abelchain1}
\end{array}\\
&&\mathbb{D}_{A}^1
\left(
\begin{array}{l}
x_1\\
x_2
\end{array}
\right)\,\,\Rightarrow
\begin{array}{l}\dot{x}_1+(a_1x_1^2+a_2x_2^2)x_1=0,\\
\dot{x}_2+(a_1x_1^2+a_2x_2^2)x_2=0.
\end{array}\\
&&\mathbb{D}_{A}^2
\left(
\begin{array}{l}
x_1\\
x_2
\end{array}
\right)\,\,\Rightarrow
\begin{array}{l}
\ddot{x}_1+(2(a_1x_1^2+a_2x_2^2))\dot{x}_1
+x_1((a_1x_1^2+a_2x_2^2)^2+2a_1x_1\dot{x}_1+2a_2x_2\dot{x}_2)=0,\\
\ddot{x}_2+(2(a_1x_1^2+a_2x_2^2))\dot{x}_2
+x_2((a_1x_1^2+a_2x_2^2)^2+2a_1x_1\dot{x}_1+2a_2x_2\dot{x}_2)=0.
\end{array}\label{dvp}\\
&&\mathbb{D}_{A}^3
\left(
\begin{array}{l}
x_1\\
x_2
\end{array}
\right)\,\,\Rightarrow
\begin{array}{l}
\dddot{x}_1+3(a_1x_1^2+a_2x_2^2)\ddot{x}_1+6(a_1x_1\dot{x}_1+a_2x_2\dot{x}_2)(\dot{x}_1+x_1(a_1x_1^2+a_2x_2^2))\\
+2x_1(a_1\ddot{x}_1+a_2x_2\ddot{x}_2)+2a_1x_1\dot{x}_1^2+3(a_1x_1^2+a_2x_2^2)^2+x_1(a_1x_1^2+a_2x_2^2)^3=0,\\
\dddot{x}_2+3(a_1x_1^2+a_2x_2^2)\ddot{x}_2+6(a_1x_1\dot{x}_1+a_2x_2\dot{x}_2)(\dot{x}_2+x_2(a_1x_1^2+a_2x_2^2))\\
2x_2(a_1\ddot{x}_1+a_2x_2\ddot{x}_2)+2a_2x_2\dot{x}_2^2+3(a_1x_1^2+a_2x_2^2)^2+x_2(a_1x_1^2+a_2x_2^2)^3=0,
\end{array}\label{abelchain4}
\end{eqnarray}
and so on, where
$\mathbb{D}_{A}^l=\left(\frac{d}{dt}+(a_1x_1^2+a_2x_2^2)\right)^l$. The above integrable chains are  generalized coupled version of the Riccati and Abel chains\cite{Carinena:09}. Some of the interesting equations in the above chains are the coupled modified Emden equations (\ref{cmee}), coupled generalization of Chazy type equation\cite{Chazy:11} (\ref{chazy}) and the coupled generalized Duffing-van der Pol oscillator equations\cite{Gonz:1983} (\ref{dvp}) and so on. One can also identify a third type of integrable chain which is given as
\begin{eqnarray}
&&\mathbb{D}_N^0 
\left(
\begin{array}{l}
x_1\\
x_2
\end{array}
\right)\Rightarrow
\begin{array}{l}
x_1=0,\\
x_2=0.
\end{array}\label{newchain1}\\
&&\mathbb{D}_N^1
\left(
\begin{array}{l}
x_1\\
x_2
\end{array}
\right)\Rightarrow
\begin{array}{l}
\dot{x}_1+a_1x_1^2x_2=0,\\
\dot{x}_2+a_1x_1x_2^2=0.
\end{array}\\
&&\mathbb{D}_N^2
\left(
\begin{array}{l}
x_1\\
x_2
\end{array}
\right)\Rightarrow
\begin{array}{l}
\ddot{x}_1+2a_1x_1x_2\dot{x}_1+a_1x_1^3x_2^2+a_1(\dot{x}_1x_2+x_1\dot{x}_2)=0,\\
\ddot{x}_2+2a_1x_1x_2\dot{x}_2+a_1x_1^2x_2^3+a_1(\dot{x}_1x_2+x_1\dot{x}_2)=0.
\end{array}\label{newchain3}\\
&&\mathbb{D}_N^3
\left(
\begin{array}{l}
x_1\\
x_2
\end{array}
\right)\Rightarrow
\begin{array}{l}
\dddot{x}_1+2a_1(x_1x_2\ddot{x}_1+x_1\dot{x}_1\dot{x}_2+\dot{x}_1^2x_2)
+a_1(3x_1^2x_2\dot{x}_1+x_1^3\dot{x}_2)\\
\qquad\qquad\quad+a_1(\ddot{x}_1x_2+x_1\ddot{x}_2+2\dot{x}_1\dot{x}_2)=0,\\
\dddot{x}_2+2a_1(x_1x_2\ddot{x}_2+x_2\dot{x}_1\dot{x}_2+\dot{x}_1x_2^2)
+a_1(3x_1x_2^2\dot{x}_2+x_1\dot{x}_2^3)\\
\qquad\qquad\quad+a_1(\ddot{x}_1x_2+x_1\ddot{x}_2+2\dot{x}_1\dot{x}_2)=0,
\end{array}\label{newchain4}
\end{eqnarray}
and so on, where $\mathbb{D}_N^{(l)}=\left(\frac{d}{dt}+a_1x_1x_2\right)^l$. 
\section{Conclusion}
Li\'enard type nonlinear oscillators and their coupled versions are of much interest in science and technology 
in recent times due to their wide applicability. In this context, we have presented a brief overview of some of the recent progress made in identifying and generating integrable Li\'enard type nonlinear dynamical systems. They exhibit interesting oscillatory solutions and other properties, including quantum aspects. For example Mathews-Lakshmanan oscillators admit amplitude dependent oscillatory property and correspond to velocity dependent or position dependent mass Hamiltonian systems, while modified Emden equations possesses amplitude independent oscillatory property (isochronous oscillation) with maximal number of Lie point symmetries, {\it PT} symmetric property and so on. We have also shown that the Li\'enard type II systems admit nonstandard Lagrangian and Hamiltonian formulations. We have briefly presented a system of completely integrable $N$-coupled Li\'enard type II nonlinear oscillators.  In general, the system admits $N$ time-independent and $N$ time-dependent integrals.  We have also given a method of identifying integrable coupled nonlinear ODEs of any order from linear uncoupled ODEs of the same order by introducing suitable nonlocal transformations.  We found that the problem of solving these classes of coupled nonlinear ODEs of any order effectively reduces to solving a single first order nonlinear ODE. It is clear that more general transformations and linearizations can give rise to very many interesting new dynamical systems.

However, all the integrable systems discussed in this article have elemantary type (oscillatory) solutions.
As we have noted in the introduction there are integrable system that possess other type of solutions like elliptic functions, hyperelliptic functions, Painlev\'e transcendental functions, etc. For example special parametric choices of the coupled generalized \'Henon-Heiles\cite{Henon:69,Lakshmanan:93,Conte:05} or coupled quartic anharmonic oscillator system\cite{Bountis:82,Lakshmanan:93,Conte:05} admits elliptic function solutions and generalizations. Painlev\'e transcendental equations\cite{Painleve:06,Conte},
\begin{eqnarray}
\label{pain1}
w''&=&6w^2+z, \\
w''&=&2w^3+zw+\alpha,\\
w''&=&\frac{(w')^2}{w}-\frac{w'}{z}+\frac{\alpha w^2+\beta}{z}+\gamma w^3+\frac{\delta}{w}\\
w''&=&\frac{	(w')^2}{2w}+\frac{3}{2}w^3+4zw^2+2(z^2-\alpha)w+
\frac{\beta}{w}\\
w''&=&\bigg(\frac{1}{2w}+\frac{1}{w-1}\bigg)(w')^2-\frac{w'}{z}+\frac{(w-1)^2}{z^2}\bigg(
\alpha w+\frac{\beta}{w}\bigg)\nonumber\\
&+&\frac{\gamma w}{z}+\frac{\delta w(w+1)}{w-1}\\
w''&=&\frac{1}{2}\bigg(\frac{1}{w}+\frac{1}{w-1}+\frac{1}{w-z}\bigg)(w')^2-\bigg(\frac{1}{z}+\frac{1}{z-1}+\frac{1}{w-z}\bigg)w'\nonumber\\
&+&\frac{w(w-1)(w-z)}{z^2(z-1)^2}\bigg[\alpha+
\frac{\beta z}{w^2}\bigg)+\frac{\gamma (z-1)}{(w-1)^2}+\frac{\delta z(z-1)}{(w-z)^2}\bigg],
\end{eqnarray}
require the introduction of new transcendental functions to solve them\cite{Ince:56}. These equations are linearizable in a more generalized sense. Coupled versions of such systems and their linearization properties are all challenging future problems.  One can employ several recently developed methods for this purpose, for example modified Prelle-Singer method\cite{Chandrasekar:05c} , generalized linearization procedure\cite{Chandrasekar:06}, method of Jacobic multipliers\cite{Guha:11}, Darboux polynomial method\cite{Darboux}, factorization method\cite{Reyes:05,Reyes:08,Hazra:12}, inverse scattering transform method\cite{Ablowitz:91} and so on. One can expect continued multifaceted progress in these topics.

\section*{Acknowledgments}
ML wishes to thank Professor P. M. Mathews for his inspiring guidance during the early stages of this work in 1970s. Both the authors record their appreciation of their current collaborators Dr. M. Senthilvelan and Dr. Gladwin Pradeep, as well other younger colleagues at Bharathidasan University.
The work is supported by the Department of Science and Technology (DST)--Ramanna program and DST--IRHPA research project. ML is also supported by a DAE Raja Ramanna Fellowship. 
\appendix
\section*{Appendix A : Some technical terminologies}
In this Appendix, we briefly explain certain technical terms used in this article.  

\begin{enumerate}
\item {\bf Integrable system}

A dynamical system is called integrable typically if the underlying nonlinear differential equation admits sufficient number of independent integrals of motion so that the equation of motion can in principle be integrated in terms of regular (meromorphic) functions.

\item {\bf Lienard equation}

A second order equation of the form $\ddot{x}+f(x)\dot{x}+g(x)=0,$ where $f(x),\,g(x)$ are continuously differentiable functions on the real line, $f$ is an even function and $g$ is an odd function, is usually called  the Li\'enard equation.  It has been studied in the context of vacuum tube oscillating circuits. In general Li\'enard equation has a unique and stable limit cycle solution if it satisfies the following conditions:

\begin{enumerate}
\item $g(x)>0$ for $x>0$,  
\item $\displaystyle F(x)=\int_0^xf(x')dx'$ has exactly one positive root at some value $m$, where $F(x)<0$ for $0<x<m$ and $F(x)>0$ and monotonic for $x>p$ and
\item $lim_{x\rightarrow\infty}F(x)=\infty$.
\end{enumerate}
A typical example of Li\'enard equation is the well known van der Pol equation, $\ddot{x}-\mu(1-x^2)\dot{x}+x=0$.
As a generalization to the Li\'enard equation one can include an additional quadratic term of $\dot{x}$ such that the equation takes the form
of Eq. (9) with a redesignation of the functions $f(x),\,g(x)$ and $h(x)$.

\item {\bf Emden equation}

The Lane-Emden equation arises in the study of the gravitational potential of a Newtonian self-gravitating spherically symmetric, polytropic fluid and is of the form $\ddot{u}+\frac{2}{t}\dot{u}+u^n=0.$

Using a  series of transformations one can transform the above equation to the form\cite{dixon}
$\ddot{x}+\alpha x \dot{x}+\beta x^3=0.$  We call this equation with an additional linear force $\omega_0^2x$ as the modified Emden equation, see Eq. (49).

\item {\bf \emph{PT} symmetry}

A dynamical equation which is invariant under the combined transformation $x\rightarrow-x$ and $t\rightarrow-t$ is known to be \emph{PT} symmetric.  \emph{PT} symmetric Hamiltonians  are important in the study of non-Hermitian quantum mechanics, where one requires additionally $i\rightarrow -i$. Here the energy eigenvalues can be real inspite of the Hamiltonian being non-Hermitian\cite{bender_r}.

\item {\bf Isochronous oscillators}

An oscillator whose frequency of oscillation is independent of the amplitude is called an isochronous oscillator.  A simple example is the linear harmonic oscillator.  The interesting fact is that even nonlinear oscillators of suitable forms can exhibit isochronous oscillations\cite{Calogero:08,Gladwin Pradeep:09b,Chandrasekar:12}.

\item {\bf Nonstandard Lagrangian/Hamiltonian}

Standard Lagrangian/natural Lagrangian is written as the difference between kinetic energy and potential energy.  However, there are situations where one is unable to write an identified Lagrangian as above.  Such Lagrangians are called nonstandard Lagrangians.  The corresponding Hamiltonians, which cannot be written as the sum of kinetic and potential energy terms, are called nonstandard Hamiltonians\cite{Gladwin Pradeep:09a,Musielak:08}.

\item {\bf Singular Lagrangian/Hamiltonian}

A given Lagrangian is known as a singular/degenerate Lagrangian\cite{singular2} if the determinant of the Hessian matrix is zero, that is the condition
$\displaystyle\vline\frac{\partial^2L}{\partial \dot{x}_i\partial \dot{x}_j}\vline=0,\quad i,j=1,2,\ldots,N,$
is valid.  Similarly, a given Hamiltonian is known as a singular/degenerate Hamiltonian if the determinant of the Hessian matrix is zero, that is the condition $\displaystyle\vline\frac{\partial^2H}{\partial p_i\partial p_j}\vline=0,\quad i,j=1,2,\ldots,N,$
is satisfied.

\end{enumerate}


\begin{thebibliography}{10}

\bibitem{Guc:1983}
J. Guckenheimer and P. Holmes, Nonlinear oscillations, dynamical 
systems and bifurcations of Vector Fields, Springer-Verlag, New York (1983).

\bibitem{Tab:1989}
M. Tabor, Chaos and integrability in nonlinear dynamics: An introduction,
John Wiley \& Sonc. Inc, New York (1989).

\bibitem{Nayfeh:95}
A. H. Nayfeh and D. T. Mook, Nonlinear oscillations, John Wiley Sons, New York (1995).

\bibitem{Wig:2003}
S. Wiggins, Introduction to Applied Nonlinear Dynamical Systems 
and Chaos, Springer-Verlag, New York (2003).

\bibitem{Lakshmanan:03}
M. Lakshmanan and S. Rajasekar, Nonlinear dynamics: Integrability
chaos and patterns, Springer-Verlag,  New York (2003).

\bibitem{Calogero:08}
F. Calogero,  Isochronous systems, Oxford University Press, Oxford (2008).

\bibitem{Murphy:69}
G. M. Murphy , Ordinary differential equations and their solutions,
Affiliated East-west press, New Delhi (1969).

\bibitem{Mathews:74}
P. M. Mathews and M. Lakshmanan, On a unique nonlinear oscillator, Quart. Appl. Math. {\bf 32}, 215 (1974)

\bibitem{bend:98}
B. Belchev and M. A. Walton, The Morse potential and phase-space quantum mechanics
arXiv.org:1001.4816v1 (2010); D. Zhu, A new potential with the spectrum of an isotonic oscillator, J. Phys. A: Math. Gen. {\bf20}, 4331 (1987)

\bibitem{Delbourgo}
R. Delbourgo, A. Salam and J. Strathdee, Infinities of nonlinear and Lagrangian theories. Phys. Rev. {\bf 187}, 1999–2007  (1969).
\bibitem{Koc}
R. Koc ̧ and M. Koca, A systematic study on the exact solution of the position-dependent mass Schr\"odinger equation. J. Phys. A {\bf 36}, 8105–8112 (2003).
\bibitem{Gladwin Pradeep:09a}
R. Gladwin Pradeep, V. K. Chandrasekar, M. Senthilvelan and M. Lakshmanan, Nonstandard conserved Hamiltonian structures in dissipative/damped systems: Nonlinear generalizations of damped harmonic oscillator, J. Math. Phys. {\bf50}, 052901 (2009).

\bibitem{Mathews:75}
P. M. Mathews and M. Lakshmanan, A quantum mechanically solvable nonpolynomial Langrangian with velocity-dependent interaction, Nuovo Cimento {\bf A 26}, 299 (1975).

\bibitem{Lakshmanan:75}
M. Lakshmanan and K. Eswaran, Quantum dynamics of a solvable nonlinear chiral model, J. Phys.  A {\bf8}, 1658 (1975).

\bibitem{higgs}
P. W. Higgs, Dynamical symmetries in a spherical geometry I, J. Phys. A: Math. Gen. {\bf12}, 309 (1979).

\bibitem{Leemon:79}
H. I. Leemon, 
Dynamical symmetries in a spherical geometry. II,  J. Phys. A: Math. Gen. {\bf12}, 489 (1979).

\bibitem{13}
J. F. Carinena, M. F. Ranada, and M. Santander, One-dimensional model of a quantum non-linear harmonic oscillator, Rep. Math. Phys. {\bf54}, 285 (2004).

\bibitem{15}
A. Venkatesan and M. Lakshmanan, Nonlinear dynamics of damped and driven velocity-dependent systems, Phys. Rev. E {\bf55}, 5134 (1997).

\bibitem{Carinena:04}
 J. F. Carinena, M. F. Ranada, M. Santander and M. Senthilvelan, A non-linear Oscillator with quasi-Harmonic behaviour: two- and n-dimensional oscillators, Nonlinearity {\bf17}, 1941 (2004)

\bibitem{Carinena:07}
 J. F. Carinena, M. F. Ranada and M. Santander , A quantum exactly solvable non-linear oscillator with quasi-harmonic behaviour, Ann. Phys. {\bf322}, 434 (2007) 

\bibitem{Tewari:13}
A. Tewari, S. N. Pandey, M. Senthilvelan and M. Lakshmanan, Classification of Lie point symmetries for quadratic Li$\acute{\textbf{e}}$nard type equation $\ddot{x}+f(x)\dot{x}^2+g(x)=0$, arXiv:1302.0350  (2013).

\bibitem{Bhuneshwari:12}
A. Bhuvaneswari, V. K. Chandrasekar, M. Senthilvelan and M. Lakshmanan, On the complete integrability of a nonlinear oscillator from group theoretical perspective, J. Math. Phys.  {\bf53}, 073504 (2012).

\bibitem{Bagchi:13}
B. Bagchi, S. Das, S. Ghosh and S. Poria, Nonlinear dynamics of a position-dependent mass-driven 
Duffing-type oscillator,  J. Phys. A {\bf46}, 032001 (2013).

\bibitem{Cruz:13}
S. C. Cruz and O. Rosas-Ortiz, Dynamical Equations, Invariants and Spectrum Generating Algebras of Mechanical Systems with Position-Dependent Mass, SIGMA {\bf9}, 004 (2013).

\bibitem{Pandey:09}
S. N. Pandey, P. S. Bindu, M. Senthilvelan and M. Lakshmanan, A Group Theoretical Identification of Integrable Cases of the Liénard Type Equation $\ddot{x}+f(x)\dot{x}+g(x) = 0$ : Part I: Equations having Non-maximal Number of Lie point Symmetries, J. Math. Phys. {\bf 50}, 082702 (2009); A Group Theoretical Identification of Integrable Equations in the Liénard Type Equation $\ddot{x}+f(x)\dot{x}+g(x) = 0$ : Part II: Equations having Maximal Lie Point Symmetries, J. Math. Phys. {\bf 50}, 102701 (2009)

\bibitem{mah:1985}
F. M. Mahomed and P. G. L. Leach, The Lie Algebra SL (3,R) and Linearization, Quaestiones Math., 12, 121 (1989); P. G. L. Leach, M. R. Feix and S. Bouquet,  Analysis and solution of a nonlinear second-order differential equation through rescaling and through a dynamical point of view, J. Math. Phys. {\bf 29}, 
2563 (1988).

\bibitem{Chandrasekar:05}
 V. K. Chandrasekar, M. Senthilvelan and M. Lakshmanan, Unusual Liénard-type nonlinear oscillator, Phys. Rev. E , 72, 066203 (2005); On the complete integrability and linearization of certain second-order nonlinear ordinary differential equations, Proc. Roy. Soc. London  {\bf A461}, 2451 (2005).

\bibitem{Chithika Ruby:12}
V. Chithika Ruby , M. Senthilvelan and M. Lakshmanan, Exact quantization of a PT symmetric (reversible) 
Li\'enard type nonlinear oscillator, J. Phys. A : Math. Theor. {\bf 45}, 382002 (2012).

\bibitem{von}
O. von Roos, Position-dependent effective masses in semiconductor theory, Phys. Rev. B {\bf 27}, 7547 
(1983); 
O. von Roos  and H. Mavromatis, Position-dependent effective masses in semiconductor theory. II, Phys. Rev. B {\bf 31}, 2294 (1985).

\bibitem{bender_r}
C. M. Bender and D. W. Hook, Quantum tunneling as a classical anomaly, J. Phys. A: Math. Theor. {\bf 44}, 372001 (2011);
C. M. Bender, Making sense of non-Hermitian Hamiltonians, Rep. Prog. Phys. {\bf 70}, 947 (2007).

\bibitem{feix:1997}
M. R. Feix, C. Geronimi, L. Cairo, P. G. L. Leach, R. L. Lemmer and S. Bouquet, On the singularity analysis of ordinary differential equations invariant under time translation and rescaling,  
J. Phys. A: Math. Gen. {\bf 30}, 7437 (1997).

\bibitem{Chandrasekar:07}
V. K. Chandrasekar, M. Senthilvelan, M. Lakshmanan, On the general solution for the modified Emden-type 
equation $\ddot{x}+\alpha x\dot{x}+\beta x^3=0$, J. Phys. A : Math. Gen. {\bf 40}, 4717 (2007)

\bibitem{Calogero:07}
F. Calogero  and F. Leyvraz,  General technique to produce isochronous Hamiltonian, J. Phys. A: Math. Theor. {\bf40}, 12931 (2007).

\bibitem{Calogero:08a}
F. Calogero  and F. Leyvraz,  Examples of isochronous Hamiltonians: classical and quantal treatments,
J. Phys. A: Math. Theor. {\bf41},  175202 (2008).

\bibitem{singular2}
E. C. G. Sudarshan and N. Mukunda, Classical Dynamics : A Modern Perspective , John Wiley \& Sons, New York (1974).


\bibitem{Gladwin Pradeep:09b}
R. Gladwin Pradeep, V. K. Chandrasekar, M. Senthilvelan and M. Lakshmanan, Dynamics of a completely integrable N-coupled Li\'enard-type nonlinear oscillator, J. Phys. A Math. Theor.  {\bf 42}, 135206 (2009).

\bibitem{Musielak:08}
Z. E. Musielak, Standard and non-standard Lagrangians for dissipative dynamical systems with variable coefficients, J. Phys. A: Math. Theor. {\bf41}, 055205 (2008).

\bibitem{Cie:10}
Jan L. Cie\'sli\'nski and T. Nikiciuk, A direct approach to the construction of standard and non-standard Lagrangians for dissipative-like dynamical systems with variable coefficients,  J. Phys. A: Math. Theor. {\bf43}, 175205 (2010).

\bibitem{Chandrasekar:12}
V. K. Chandrasekar, Jane H. Sheeba, R. Gladwin Pradeep, R. S. Divyashree and M. Lakshmanan,  A class of solvable coupled nonlinear oscillators with amplitude independent frequencies, Phys. Lett. A  {\bf 376}, 2188 (2012).

\bibitem{Gladwin Pradeep:12c}
A. Durga Devi, R Gladwin Pradeep, V K Chandrasekar and M Lakshmanan, Method of generating $N$-dimensional isochronous nonsingular Hamiltonian systems, J. Nonlinear Math. Phys. (Accepted for publication); arXiv:1207.4611 (2013)

\bibitem{Chandrasekar:06}
V. K. Chandrasekar, M. Senthilvelan , A. Kundu and M. Lakshmanan,  A nonlocal connection between certain linear and nonlinear ordinary differential equations/oscillators, 
J. Phys. A : Math. Gen. {\bf39}, 9743 (2006).

\bibitem{Gladwin Pradeep:10}
R. Gladwin Pradeep, V. K. Chandrasekar, M. Senthilvelan and M. Lakshmanan, A nonlocal connection between certain linear and nonlinear ordinary differential equations: Extension to coupled equations, J. Math. Phys. {\bf 51}, 103513 (2010).

\bibitem{Gambier:10}
B. Gambier, Sur les equations differ entielles du second ordre et du premier degre dont I'integrale generate est a points critiques fixes, Acta Math. {\bf33}, 1 (1910).

\bibitem{GR}
B. Grammaticos, A. Ramani, The Gambier mapping, Phys.~A {\bf 223}, 125 (1996).

\bibitem{GRL}
B. Grammaticos, A. Ramani, S. Lafortune, The Gambier mapping, revisted, Phys.~A {\bf 253}, 260 (1998).

\bibitem{Guga1}
P. Guha, A. G. Choudhury and B. Grammaticos, Dynamical Studies of Equations from the Gambier Family, SIGMA {\bf7}, 028 (2011).

\bibitem{ref10}
I. S. Gradshteyn and I. M. Ryzhik, Table of Integrals, Series and 
Products, Academic press, London (1980).

\bibitem{Carinena:09}
J. F. Carinena, P. Guha and M. F. Ranada, Higher-order Abel equations: Lagrangian formalism, first integrals and Darboux polynomials, Nonlinearity {\bf22}, 2953 (2009).

\bibitem{Chazy:11}
J. Chazy, Sur la limitation du degr\'e des coëfficients des\'equations diff\'erentielles alg\'ebriques \'a points critiques fixes, Acta Math. {\bf34}, 317 (1911).

\bibitem{Gonz:1983}
D. L. Gonzalez and O. Piro, Chaos in a nonlinear driven oscillator with exact solution, Phys. Rev. Lett. {\bf 50}, 870 (1983); Disappearance of chaos and integrability in an externally modulated nonlinear oscillator \emph{Phys. Rev. A} {\bf 30}, 2788 (1984).

\bibitem{Henon:69}
M. \'Henon and  C. Heiles, 1964 The applicability of the third integral of motion: some numerical experiments, Astron. J. {\bf69}, 73 (1964). 

\bibitem{Lakshmanan:93}
M. Lakshmanan and R. Sahadevan, Painlev\'e analysis, Lie
symmetries, and integrability of coupled nonlinear oscillators of
polynomial type, Phys. Rep.  \,\textbf {224}, 1 (1993).


\bibitem{Conte:05}
R. Conte, M. Musette and C. Verhoeven, Completeness of the cubic and quartic Henon-Heiles Hamiltonians, 
Theoretical and Mathematical Physics, {\bf144}, 888 (2005).

\bibitem{Bountis:82}
T. Bountis, H. Segur and F. Vivaldi, Integrable Hamiltonian systems and the Painlev\'e property, Phys. Rev. A {\bf25}, 1257 (1982).


\bibitem{Painleve:06}
P. Painlev\'e, Surles $\acute{e}$́quations diff$\acute{e}$rentielles du second ordre $\acute{a}$ points critiques fix$\acute{e}$s, C. R. Acad. Sc. Paris,  {\bf143}, 1111 (1906).

\bibitem{Conte}
R. Conte,  The Painlev$\acute{e}$ property: One century 
later,  New York, Springer (1999).

\bibitem{Ince:56}
 E. L. Ince, Ordinary differential equations, Dover, NewYork (1956).

\bibitem{Chandrasekar:05c}
V. K. Chandrasekar, M. Senthilvelan and M. Lakshmanan, Extended Prelle-Singer method and integrability/solvability of a class of nonlinear nth order ordinary differential equations,  J. Nonlinear Math. Phys. {\bf 12}, 184 (2005).

\bibitem{Chandrasekar:06}
V. K. Chandrasekar, M. Senthilvelan and M. Lakshmanan, A unification in the theory of linearization of second-order nonlinear ordinary differential equations, J. Phys. A: Math. Theor. {\bf 39}, L69 (2006).

\bibitem{Guha:11}
 P. Guha and A.  G. Choudhury, The role of the Jacobi last multiplier and isochronous systems,  
Pramana {\bf 77}, 917 (2011).

\bibitem{Darboux}
G. Darboux, Memoire sur les \'equations diff’\'erentielles du premier ordre et du premier
degr\'e. Bull. Sci. Math. {\bf32}, 60-96, 123–144, 151–200 (1878).


\bibitem{Reyes:05}
O. Cornejo-P\'erez and H. C. Rosu, Nonlinear second order Ode's: Factorizations and particular solutions, Prog. Theor. Phys. {\bf 114}  533 (2005).

\bibitem{Reyes:08}
M. A. Reyes and H. C. Rosu, Riccati-parameter solutions of nonlinear second-order ODEs, J. Phys. A: Math. Theor. {\bf 41}, 285206 (2008).

\bibitem{Hazra:12}
T. Hazra, V. K. Chandrasekar, R Gladwin Pradeep and M. Lakshmanan, Exact solutions of coupled Li\'enard-type nonlinear systems using factorization technique, J. Math. Phys. {\bf 53}, 023511 (2012).

\bibitem{Ablowitz:91}
M. Ablowitz, P. Clarkson, Solitons, Nonlinear evolution equations and inverse scattering, Cambridge University Press, Cambridge (1991).

\bibitem{dixon}
 J M Dixon and J A Tuszynski, Solutions of a generalized Emden equation and their physical significance Phys. Rev. A {\bf 41} 4166–73 (1990).
\end{thebibliography}
\end{document}